\documentclass[final]{hubsarxiv}

\usepackage{longtable}
\usepackage{flafter}   
\graphicspath{{figures/}}
\newcolumntype{L}[1]{>{\raggedright\arraybackslash}p{#1}}
\makeatletter
\def\HardFloatBarrier{\par\begingroup\let\@elt\relax\edef\@tempa{\@botlist\@deferlist\@dbldeferlist}%
 \ifx\@tempa\@empty\else\clearpage\fi\endgroup}
\def\LongtableBarrier{\par\begingroup\let\@elt\relax\edef\@tempa{\@botlist\@deferlist\@dbldeferlist}%
 \ifx\@tempa\@empty\else\newpage\fi\endgroup}
\makeatother

\title{How ready is the United States commercial building stock for artificial intelligence? A national estimate from energy codes and survey data}
\shorttitle{AI readiness of the U.S. commercial building stock}
\shortauthors{Jung}
\author{Wooyoung Jung\affmark{a,}\corrmark\orcid{0000-0003-2746-4983}}
\affil{\affmark{a}Department of Civil and Architectural Engineering and Mechanics, University of Arizona, Tucson, AZ, USA}
\corresponding{wooyoung@arizona.edu}
\paperno{2026-011}
\paperdate{November 2026}
\versionnote{v1 (2026-11-15): first release.}
\license{CC BY-NC-ND 4.0}
\keywords{AI readiness; Commercial building stock; Building energy codes; Multiple imputation; Building data; CBECS}
\abstract{Buildings serve as platforms for artificial intelligence (AI) through the data they generate, but
the building stock's readiness as an AI platform has not been assessed. A recent framework measures
a building's AI readiness as the share of 76 answer levels, across 31 types of
operators' questions, that its recorded data could reach, and has so far been applied only to
individual buildings. This study applies it to the 2018 U.S. commercial building stock. It combines
6,436 records of the Commercial Buildings Energy Consumption Survey with ASHRAE 90.1's data
requirements, state code adoption, advanced-meter statistics and 45 buildings' knowledge
graphs, imputing each building's code edition and data recording. Weighted by floor area, the stock
could answer on average 0.145 (95\% interval 0.127--0.163) of these answer levels. For
92.0\% of floor area its data answer less than half of the questions about current status,
and most floor area stays below that mark under every tested assumption about compliance,
description and the reference buildings. More than half of what
installed equipment could answer is lost in recording, mostly because conventional thermostats store
nothing and meters lack interval readings. Readiness rises with size, from 0.073 in
buildings of at most 5,000~ft\textsuperscript{2} to 0.173 at 25,000~ft\textsuperscript{2}
or more. Only 1.6\% of floor area was built under a code edition requiring recording,
introduced in 90.1-2013 for large buildings. The study gives the first national baseline of
commercial-building AI readiness and a public-data method that can be reapplied to each new survey
to track the stock.}

\begin{document}
\maketitle
\section{Introduction}
\label{sec:intro}

Efforts to cut the energy use of commercial buildings, 6.8 quadrillion British thermal units in the
United States in 2018 \citep{eia2022cbecspress}, have steadily added data to buildings through
energy codes, owners' investments in automation and utilities' advanced meters
\citep{athalye2016impacts,trenbath2022sensors,eia2023ami}. These data make buildings potential
platforms for artificial intelligence (AI). Analytics built on them already save energy, and owners
in a public-private campaign covering more than 400 million square feet who reported results
achieved median annual cost savings of 7\% \citep{kramer2019analytics}. AI research has gone
further, from diagnosing equipment faults and scheduling equipment to answering operators' and
occupants' questions
\citep{zhao2019aifdd,jung2026builddx,wang2020rl,mulayim2024llm,jung2026hemseval,jung2026collab}. Its
latest step is AI agents built on large language models, which answer operators' questions over a
building's sensor data and turn their instructions into control actions
\citep{lee2025hvacqa,liu2026intent}.

The promise of AI for buildings is clear, but the status of buildings as AI platforms has received
less attention. Buildings stay in service for decades, and most of the existing stock was built
before AI was a practical tool. They have acquired and stored data only where forced by code or
justified by owners. In 2012, centralized automation was far less common in buildings under
50,000~ft\textsuperscript{2} than in larger ones \citep{doe2020sensors}. Such differences decide
what an AI agent can do in a building, because an agent can answer an operator's question only from
data that the building records and links to the equipment or space they measure. Existing
assessments of whether a building is ready for AI, such as the European smart readiness indicator
\citep{eu2020sri} and WiredScore's SmartScore \citep{wiredscore2026}, rate the services and
technology a building has installed. Neither bases its score on the data the building records, so
neither shows which of an operator's questions an agent could answer from those data
\citep{jung2026readiness}.

The AI readiness framework \citep{jung2026readiness} was proposed to fill this gap, with AI agents
in view. It fixes 31 types of question that operators ask about a building's status, the
summaries of its performance and the diagnosis of its problems, and scores a building by the share
of them its data could answer. That share is a ceiling on what any agent can answer from the
building, and the framework measures separately how much of it a reference agent realizes. For
example, a question about whether a zone is too warm can be answered only if the building records
zone temperature and links it to the zone it describes. However, the framework's assessment of AI
readiness has so far been applied only to a small set of buildings, and the landscape of AI
readiness across the national stock is unknown.

This study aims to map the AI readiness landscape for the United States (U.S.) commercial building
stock. The U.S. Energy Information Administration (EIA) offers a national survey that makes such a
map possible. Its Commercial Buildings Energy Consumption Survey (CBECS) represents the commercial
stock and describes each building's size, type, age, energy use and reported
controls~\citep{eia2023cbecsguide}. CBECS records two kinds of information that indicate a
building's AI readiness. The first points to the energy code. A building's construction period and
census division indicate which edition of the American Society of Heating, Refrigerating and
Air-Conditioning Engineers (ASHRAE) Standard 90.1, the model standard behind U.S. commercial energy
codes, was likely in force when it was built or renovated, and that edition fixes the sensors and
meters a compliant building had to install. The second is the set of controls a building reports,
such as a building automation system (BAS) or lighting occupancy sensors. Together they let the
study estimate what data streams a building is likely to have, and therefore what share of those
questions an AI agent could at most answer. The study addresses three research questions.
\begin{itemize}
  \item \textbf{RQ1, level.} On average, weighted by floor area, what share of those operator questions
    can the stock's data answer, and how much floor area can answer at least half of the questions about
    status, summaries or diagnosis? How wide a range do public data allow?
  \item \textbf{RQ2, sources and distribution.} Which kinds of data supply that AI readiness, and
    how does it vary with building size, type and age?
  \item \textbf{RQ3, the energy code.} Has the energy code raised the AI readiness of the existing
    stock, and what would its monitoring requirement add where it applies?
\end{itemize}
Three hypotheses, each drawn from how buildings have acquired their data, guide the analysis:
\begin{itemize}
  \item \textbf{H1 (RQ1).} Because older editions of the energy code required little more than a
    thermostat and a utility meter, most floor area would answer less than half of what the status
    questions ask (e.g., whether a space is occupied, whether its lights are on, or what the CO$_2$
    level is).
  \item \textbf{H2 (RQ2).} Because automation systems have been adopted mostly by large buildings
    \citep{katipamula2012small}, AI readiness would rise with building size, and large buildings
    would hold a larger share of the floor area that reaches the status tier than of the stock's
    buildings.
  \item \textbf{H3 (RQ3).} Because the energy code's one requirement to record data entered with its
    2013 edition and binds only large new buildings, the code would explain little of the
    stock's AI readiness: almost all floor area would sit at the thermostat-and-meter profile (zone
    temperature, its setpoint and a whole-building meter), and the code-required level would barely
    change across editions before 90.1-2013.
\end{itemize}

To answer these questions, the study combines the 6,436 public records of the 2018
CBECS \citep{eia2022cbecsmicrodata} with the requirements of successive 90.1 editions and the
framework's question set, whose 31 question types can each be answered at up to three
answer levels, from a value alone to a value with its cause, 76 answer levels in all. Each
survey clue takes the estimate only part of the way. Period and division narrow the code edition
without fixing it, because codes are adopted state by state and the public file gives neither the
state nor the year of construction. A reported control shows that the equipment is there but says
nothing about whether its data are recorded over time. The study therefore imputes the edition and
the recording of each stream repeatedly from public evidence, a procedure known as multiple
imputation, and combines the results across imputations \citep{rubin1987multiple}. Each building
receives three values, each the share of the 76 answer levels that a set of its streams reaches: a
code-required level from the streams its code edition required, a declared level that adds the
streams its reported controls imply, and a linked level that keeps only the streams that are
recorded and described. Two premises that no public data inform, compliance with the code and the
description of recorded data, are first set in the buildings' favor and then lowered across their
whole range, to show whether any conclusion depends on them. The inputs are checked against
published values, among them the survey's stock totals, its worked standard-error example and the
framework's published scores.

This study gives a first stock-level estimate of the AI readiness of U.S. commercial buildings,
conditional on the framework's question set, with intervals that carry sampling and imputation
uncertainty, and shows where building data are lost between installation and use. Its tables of code
requirements, reported controls and recording routes form a reusable public-data method that can be
applied to the next national survey wherever its questions match those of 2018. %

\section{Literature review}
\label{sec:related}

Estimating the AI readiness of a building stock draws on four bodies of work: what the energy code
required each building to install, what data buildings actually keep, what AI applications need
from those data, and how AI readiness has been measured. This section reviews each in turn.

\subsection{What the energy code requires and how it reaches the stock}
\label{sec:rel-code}

Energy codes have expanded the sensing and controls that new buildings must install, mostly as a
means to energy savings. The determinations of the U.S. Department of Energy (DOE) estimate what
each new edition of ASHRAE Standard 90.1 saves over its predecessor~\citep{doe2008determination,
doe2011determination2007,doe2011determination2010, doe2014determination,doe2018determination}, and
state adoption analyses track which edition each state has adopted and translate editions of the
International Energy Conservation Code (IECC) into their 90.1 equivalents
\citep{belzer2010retrospective, athalye2016impacts,tyler2021impacts}.

Each edition has added sensing and controls, and about 30\% of the requirements added to the model
codes since 2004 concern controls \citep{rosenberg2017controls}. Over these editions, the code has
moved from controls that act on what they sense toward data that can be kept and analyzed. Standard
90.1-2004 required lighting in classrooms, conference rooms and break rooms to switch off
automatically after the occupants leave~\citep{halverson2006comparison}, a control that responds on
the spot, and 90.1-2010 extended automatic shutoff to more space
types~\citep{halverson2011qualitative}. 90.1-2013 introduced the monitoring of a building's
electricity use by end use for new buildings of at least 25,000~ft\textsuperscript{2}, with readings
recorded every 15 minutes and kept for 36 months, along with whole-building
monitoring of gas, steam and chilled water \citep{halverson2014qualitative}. 90.1-2016 extended
lighting controls to alterations of existing buildings and added fault detection, with permanently
installed temperature sensors, to direct-expansion cooling units with economizers
\citep{doe2017savings2016}. Recorded data of this kind are what analytics and AI applications draw
on.

\subsection{What data buildings keep}
\label{sec:rel-data}

The code fixes what a building must install, but whether its data are kept depends largely on
choices the code leaves to owners and utilities. Buildings keep data through three main routes: an
automation system's trend records, connected thermostats and utility meters.

An automation system can keep trend records of the points it reads and controls
\citep{granderson2016eis}, but automation has spread slowly. Energy management and control systems
served about 33\% of commercial floor area in 1999, in about a tenth of buildings
\citep{brambley2005sensors}, and centralized automation covered 22\% of floor area in 2003 and 43\%
in 2012 \citep{doe2020sensors}. Only 8\% of small commercial buildings had installed optimized
sensors and controls, largely because of cost, although such systems could save 29\% of annual
energy in commercial buildings \citep{trenbath2022sensors}. Where an automation system exists, its
data are not always usable. Analytics programs report that data from older systems are hard to
extract and that inconsistent point names make integrating them labor-intensive
\citep{kramer2020business}.

A connected thermostat keeps the record that a conventional thermostat discards. It controls the
heating and cooling of its zone like any thermostat but is linked to the internet, and models
certified under ENERGY STAR must give users information on heating, ventilation and air-conditioning
(HVAC) energy use, such as monthly run time \citep{energystar2026thermostats}. Records that owners
have shared show what such devices hold: indoor and outdoor temperatures, occupant preferences and
HVAC usage \citep{ecobee2026dyd}. Through ecobee's Donate Your Data program, customers shared data
from more than 10,000 connected thermostats across North America over several years, which removed
the small samples and short observation periods that had limited earlier studies of thermostat
behavior \citep{huchuk2018longitudinal}, and the same data later supported models that predict home
occupancy \citep{huchuk2019occupancy} and representative U.S. residential occupancy schedules
\citep{jung2023occupancy}. In commercial buildings, an estimated 5\% used internet-connected
thermostats in 2018 \citep{eia2021cbecspress}. In small commercial buildings, which seldom have an
automation system, connected thermostats offer a low-cost way to collect data across a portfolio,
and data from more than 620 thermostats in 250 such buildings in Ontario showed how occupants
override setpoints \citep{huchuk2021commercial}.

A utility meter is on every electrified building, and whether it keeps readings at intervals depends
on its type. Advanced meters, which record use at least hourly and share the data with the utility
and the customer at least daily, made up 54.0\% of commercial meters in 2018
\citep{eia2018form861} and 69\% in 2021 \citep{eia2023ami}, and in small buildings their interval
data can partly stand in for automation data \citep{frank2018barriers}. Federal buildings must use,
to the maximum extent practicable, advanced meters that measure electricity and water consumption at
least hourly and provide data at least daily \citep{usc8253e}. For other buildings, benchmarking and
disclosure ordinances move annual energy use from the meter to a regulator
\citep{palmer2016benchmarking}, in 24 U.S. state and local jurisdictions by the end of 2016
\citep{mims2017benchmarking}, and rely on utilities supplying aggregated monthly consumption, with
aggregation thresholds that limit which buildings can report \citep{livingston2018meter}. Where
meter data have been shared, collections show what they can support. The Building Data Genome
Project 2 assembled two years of measured hourly data from 3,053 energy meters in 1,636
non-residential buildings at 19 sites in North America and Europe, all from sites that chose to
share \citep{miller2020bdg2}. Across the three routes, whether a building's data are kept depends on
the equipment its owner chose, the meter its utility installed and how its systems were integrated.

\subsection{What AI applications need from building data}
\label{sec:rel-needs}

Kept data serve AI only if they match what an application needs and are described well enough to be
found. AI applications draw on different combinations of data streams. Energy management and
information systems analyze interval meters and the trends that automation systems record
\citep{granderson2016eis,kramer2019analytics}. Fault detection and diagnosis needs sensor readings,
commands and labeled faults from HVAC equipment \citep{zhao2019aifdd,jung2026builddx}.
Occupant-centric HVAC operation needs occupancy data, from motion sensors, connected thermostats or
infrared arrays \citep{jung2019hitl,huchuk2019occupancy,woo2025infrared}. Model predictive control
needs a model of the building, and reinforcement learning needs extended interaction with the
building or a simulator of it \citep{drgona2020mpc,wang2020rl}.

Description is the second requirement. An application written for one building runs on another only
if both describe their points in a common way. Semantic schemas such as Brick
\citep{balaji2016brick,balaji2018brick} and the proposed ASHRAE Standard 223P
\citep{doe2026ashrae223p} provide that description as a knowledge graph of equipment, sensing points
and their relations, and language models query such knowledge graphs to answer operators' questions
in plain language \citep{mulayim2024llm,jung2026bricktrace}. Language-model agents go further,
retrieving a building's sensor data to answer its operators' questions \citep{lee2025hvacqa},
running workflows for building energy analysis \citep{zhang2025agentschema} and invoking
building-management actions from operators' instructions \citep{liu2026intent}. Methods exist to
construct such knowledge graphs automatically \citep{bhattacharya2015automated}, and the Mortar
testbed collects knowledge graphs of real buildings with their time series \citep{fierro2019mortar}.
Such descriptions remain scarce. Real knowledge graphs are rarely public, which is why synthetic
ones have been generated to stand in for them \citep{jung2026buildgraph}. Data that AI can use are
therefore data that are both recorded and described.

\subsection{How AI readiness has been measured}
\label{sec:rel-readiness}

Readiness has been measured in four ways. The most general rates a dataset: \citet{lawrence2017data}
proposed data readiness levels as a common language for how far a dataset is from use, because poor
collection practices, missing values, storage, intellectual property and privacy routinely delay
analysis. Ratings of buildings score the services a building has installed. The smart readiness
indicator counts the services present or planned across nine technical domains and grades each by
its level of functionality on expert judgment, so two panels that assessed the same building had to
reconcile their results \citep{eu2020sri,vigna2020sri}. WiredScore's SmartScore certifies technology
in nine categories from documents or a site survey \citep{wiredscore2026}, and the smart readiness
indicator showed no particular correlation with energy performance class in recent high-performance
buildings \citep{garzia2024sri}. Metadata studies assess whether a building's metadata can support
given analytics applications \citep{pritoni2021metadata,duarteroa2022stronger}. Question-answering
benchmarks, by contrast, measure the AI, testing how well it translates operators' questions into
queries over a fixed building knowledge graph \citep{jung2026build2sparql}.

The readiness framework combines these views by asking what a building's data can answer for an AI
agent \citep{jung2026readiness}. In each building it observes whether a data stream exists, whether
it is linked to the equipment or space it describes, and whether it is described well enough to be
found, and it has so far been applied only to buildings whose owners or researchers built a
knowledge graph. Stock-level studies have estimated other characteristics of the commercial stock
from the national survey, from models of energy use \citep{deng2018predictive} to the hydronic
heating and cooling systems of the stock \citep{salimianrizi2022hydronic}, and bottom-up stock
models such as ComStock combine survey and other data, statistical sampling and building energy
simulation to estimate the energy use of the stock building by building \citep{parker2023comstock}.
None has estimated the data its buildings keep or what those data could answer.

\section{Methodology}
\label{sec:method}

\subsection{Study design}
\label{sec:design}

\begin{figure}
\centering
\includegraphics[width=1.0\textwidth]{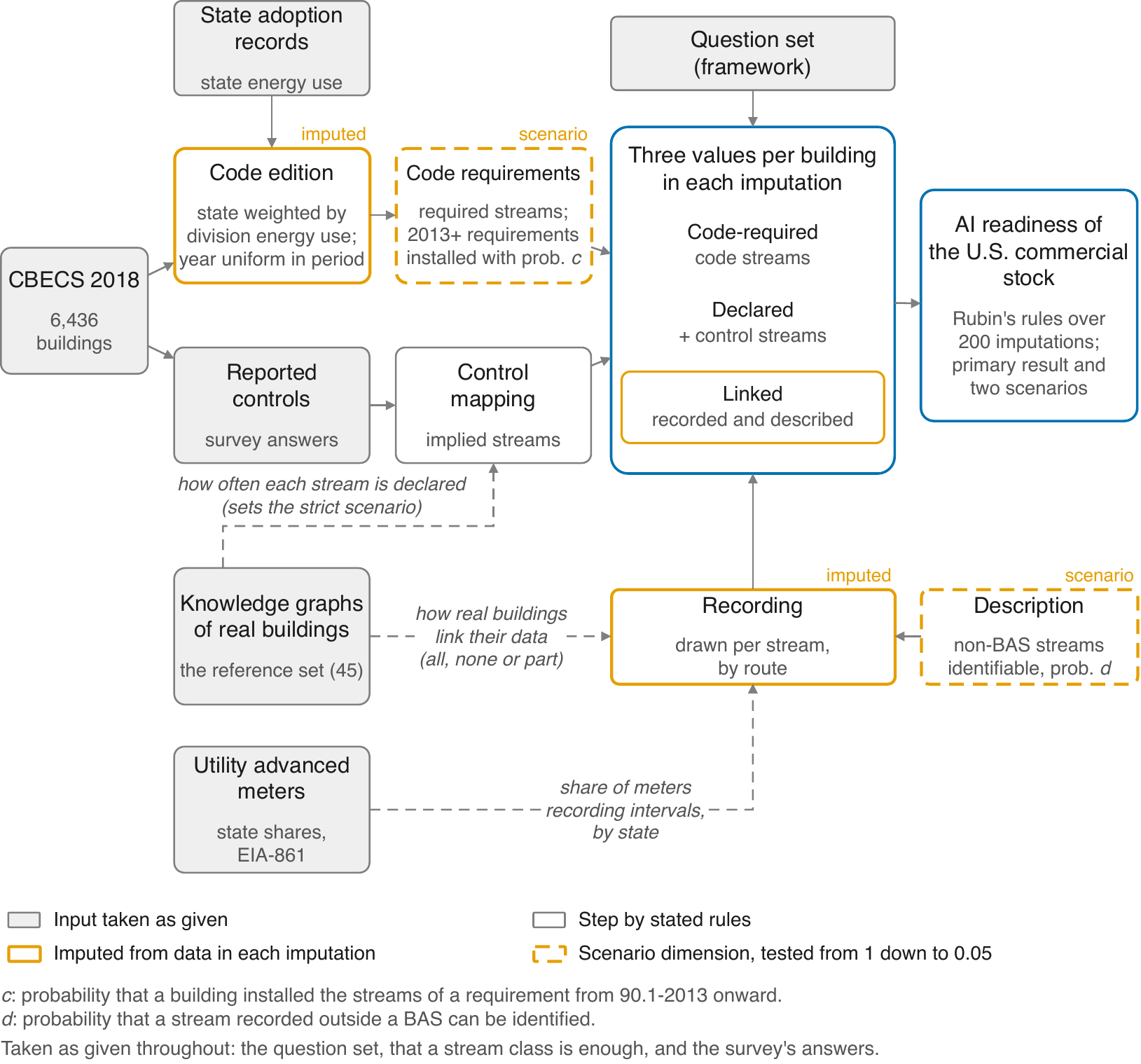}
\caption{How survey answers, code requirements, the reference set and utility meter shares combine
into three levels of AI readiness for each building.}
\label{fig:fig_method}
\end{figure}

The study estimates the AI readiness of every building in a national survey and carries the
estimates to the stock with the survey's weights (Figure~\ref{fig:fig_method}). The unit of analysis
is the survey record, and stock statements weight records by floor area unless stated otherwise.
Table~\ref{tab:design} maps each research question and hypothesis to the analyses that answer it,
and Table~\ref{tab:inputs} lists every input of the analysis, whether the survey records it, how it
is treated and how it is tested.

Each building receives three values, built up in turn. The \emph{code-required level} scores the
data streams that its code edition required. The \emph{declared level} adds the streams that its
reported controls imply. The \emph{linked level} keeps only the streams that are also recorded as
time series and described, that is, matched to the equipment or space they measure. The values are
computed in every imputation, a completed copy of the survey in which each building's code edition
and the recording of each of its streams have been drawn, and the imputations are combined with
Rubin's rules \citep{rubin1987multiple}. These rules take each estimate as the average over the
imputations and widen its interval by the spread between them, so that the interval carries both the
survey's sampling uncertainty and the uncertainty of the imputed layers (the code edition, the
recording of each stream and the control answers EIA imputed).

Because the external sources never observe the surveyed buildings, the imputation is a form of
statistical matching, or data fusion \citep{rassler2002matching,dorazio2006matching}. Its standard
assumption is conditional independence: once the variables an imputation model uses are fixed, the
imputed layer is unrelated to the building's other survey answers. Whether a BAS links its data, for
example, is assumed not to depend on the building's size, type or age. Testing the assumption would
need buildings observed in both the survey and an external source, and no such building exists. The
study therefore tests how far the least certain input, the evidence on linkage in real buildings,
would have to shift before a conclusion changes.

\begin{table}
\centering
\small\linespread{1}\selectfont
\caption{Research questions, hypotheses and analyses.}
\label{tab:design}
\setlength{\tabcolsep}{4pt}
\begin{tabular}{L{2.4cm}L{2.1cm}L{8.3cm}}
\toprule
Question & Hypotheses & Analyses \\
\midrule
RQ1, level & H1 & Mean AI readiness and the shares of floor area at the status, summary and
  diagnosis tiers, for the primary result and the strict and enriched scenarios; the share of the
  stock at or above each level of AI readiness; the same under energy-use and building-count
  weights \\
RQ2, sources and distribution & H2 & Decomposition of the mean by group of data streams;
  means and tier shares by size class, building type and construction period; decomposition of
  the variation by these factors \\
RQ3, the energy code & H3 & Code-required level by edition; share of floor area at the
  thermostat-and-meter profile; as supporting evidence, the variance that construction period and
  building type explain;
  mean and tier shares when the 90.1-2013 monitoring requirement is added where it applies and to
  every building \\
All questions & -- & Rubin's-rules intervals and the share of uncertainty due to the imputed
  layers; compliance and description scenarios; how far the reference-set evidence can be wrong
  before a conclusion changes; checks against the reference set \\
\bottomrule
\end{tabular}
\end{table}

\begin{table}
\centering
\footnotesize\linespread{1}\selectfont
\caption{Inputs and premises: whether the survey observes each, how it is treated, and how it is
tested.}
\label{tab:inputs}
\setlength{\tabcolsep}{3pt}
\begin{tabular}{L{4.2cm}L{1.15cm}L{3.7cm}L{3.8cm}}
\toprule
Information or premise & In\newline survey? & Treatment and evidence & Tested by \\
\midrule
A question is answerable only from recorded, linked streams; the question set represents what
  operators ask & -- & Taken as published (the readiness framework) & Published values reproduced;
  equal weight per domain or intent; HVAC-credit scoring \\
A class of data stream is enough; data quality, zone coverage and record length are not assessed &
  -- & Taken as given, so each value is a ceiling in these respects & -- \\
Floor area, activity, construction period, division and survey weights & Yes & Taken as recorded;
  the weights carry each building to the stock & Published totals; jackknife \\
Reported controls and renovation of each system & Yes & Taken as recorded; EIA-imputed control
  answers redrawn & Imputed records dropped \\
Energy-code edition in force & No & Imputed: state by commercial energy use in the division,
  year in the period, edition from adoption records & Equal state weights; no assumed rows; DOE
  adoption dates; editions 1 or 2 years after adoption \\
Data streams the code required & No & Read from the edition's requirements; installation is scenario
  dimension c, 0.05 to 1 & c lowered to 0.05; occupancy sensing from 90.1-2004 and from 2016
  lighting alterations; no monitoring credit in IECC states \\
Data streams the reported controls supply & No & Mapped by stated reasoning; reference-set
  frequencies & Strict scenario; the table as first written \\
Whether each stream is recorded & No & Imputed by route: reference set, taken as accurate, for
  BAS streams, the smart-thermostat answer, the state's advanced-meter share, the monitoring
  requirement & Tipping point; model checks; rare-stream prior; link-none graphs missing; monthly
  bills; smart thermostats; controls without a BAS; 2025 meter shares; daily meter access;
  large-building meters \\
Whether a recorded stream is described & No & Scenario dimension d, 0.05 to 1, for streams recorded
  outside a BAS & d lowered to 0.05 \\
\bottomrule
\end{tabular}
\end{table}

\subsection{Data}
\label{sec:data}

The data are the public microdata of the 2018 Commercial Buildings Energy Consumption
Survey (CBECS), the Energy Information Administration's (EIA) national sample of U.S. commercial
buildings \citep{eia2022cbecsmicrodata,eia2023cbecsguide}. The file holds 6,436 building
records. Each carries a final weight, which scales it to the national stock, and 151
replicate weights, from which the survey's jackknife method computes sampling intervals
\citep{eia2023cbecsguide}. Each record gives the building's floor area, principal activity,
construction period, census region and division, and the controls it reports. The survey asks
whether the building has a building automation system (BAS), recorded as an energy management and
control system, and whether that system controls heating, cooling or lighting. It also asks about
lighting occupancy sensors, smart thermostats, demand-controlled ventilation, variable-air-volume
(VAV) systems and daylight harvesting, and about recent renovation of each building system. The
smart-thermostat question is asked only of buildings without a BAS.

Two limits of the public file shape the method. Construction year is released only as 8
periods, from before 1946 to 2013--2018, and location only as 4 census regions and
9 census divisions, never as a state. Because energy codes are adopted state by state, a
building's code edition has to be inferred from its period and division. The survey records no
sensing points, and it does not ask whether a BAS keeps trend records of the points it reads or
whether any data are linked.

Four further sources supply the study's assumptions and imputation models. The requirements of
successive editions of ASHRAE Standard 90.1 and their addenda are listed with their sources in
Appendix~\ref{app:codes}. State adoption histories come from federal laboratory analyses and Department of
Energy status snapshots of 2002, 2004, 2010 and 2017
\citep{belzer2010retrospective,athalye2016impacts,tyler2021impacts,
doe2002status,doe2004status,doe2010status,doe2017status}. Utility statistics give each state's
commercial advanced-meter share in 2018 \citep{eia2018form861} and its commercial energy use
\citep{eia2026seds}.

The last is evidence on whether data in real buildings are linked. A knowledge graph of a building
is a machine-readable description of its equipment, sensors and meters, and of which of them have
recorded time series. The readiness framework assembled 45 such knowledge graphs of real
buildings, 44 from the Mortar testbed \citep{fierro2019mortar} and one from the Building TimeSeries
dataset \citep{prabowo2024bts}, and scored each one \citep{jung2026readiness}. These 45 knowledge
graphs are this study's reference set. For each building, the reference set records which data
streams the knowledge graph declares, which of them are linked to recorded series, and the AI
readiness of its declared and of its linked streams. This study uses those published results without
reprocessing the knowledge graphs, to impute whether BAS streams are recorded and to test the
reported-controls table against real buildings.

\subsection{Measuring AI readiness}
\label{sec:measure}

AI readiness is measured with the readiness framework's question set \citep{jung2026readiness}. The
set pairs six building domains (thermal, energy, air quality, equipment, occupancy and lighting)
with six operator intents (status, summary, comparison, verification, diagnosis and explanation) in
31 question types. Each type can be answered at up to three answer levels: a value alone, a
value with what it should be compared against, such as its setpoint, its own history or other zones,
and a value with its cause. Not every type has all three, so the set holds 76 answer
levels.

A building's data are described by its data-stream classes, the kinds of stream it holds, such as
zone temperature, a setpoint or a whole-building electricity meter. For each answer level the
framework lists the stream classes an answer requires, and a building reaches the level only if it
holds every listed class. A question about how much energy the building used last week, for example,
needs a whole-building meter for a value alone and, for the more complete answer, a meter on at
least one end use as well. A building's AI readiness, in the sense used throughout this paper, is
the number of answer levels its stream classes reach, divided by the 76 available.
Table~\ref{tab:app-C1} lists the question types and the stream classes each answer level requires.
The measure is a ceiling on what an AI agent could answer from the building's data. The framework
also measures how much of this ceiling an agent realizes, but this study estimates only the
ceiling, at the code-required, declared and linked levels.

\begin{table}
\centering
\caption{The question set: the data streams each answer level requires, by domain and intent.
Example
questions for every type and level are given in the readiness framework's appendix
\citep{jung2026readiness}.}
\label{tab:app-C1}
\footnotesize\linespread{1}\selectfont
\setlength{\tabcolsep}{4pt}
\begin{tabular}{L{1.6cm}L{3.0cm}L{1.8cm}L{2.2cm}L{3.4cm}}
\toprule
\textbf{Domain} & \textbf{Intents} & \textbf{Level 1} & \textbf{Level 2} & \textbf{Level 3} \\
\midrule
thermal & status, summary & T & Sp+\allowbreak{}T & — \\
thermal & comparison & T & T & ActC+\allowbreak{}ActP+\allowbreak{}Sp+\allowbreak{}T \\
thermal & verification & T & Sp+\allowbreak{}T & ActC+\allowbreak{}ActP+\allowbreak{}Sp+\allowbreak{}T \\
thermal & diagnosis, explanation & T & Sp+\allowbreak{}T & ActC+\allowbreak{}ActP+\allowbreak{}SAT+\allowbreak{}Sp+\allowbreak{}SpC+\allowbreak{}T \\
energy & status & Whole & — & — \\
energy & summary & Whole & Sub+\allowbreak{}Whole & — \\
energy & comparison, diagnosis, explanation & Whole & Whole & OAT+\allowbreak{}Sub2+\allowbreak{}Whole \\
airquality & status, summary & CO$_2$ & CO$_2$+\allowbreak{}Occ & — \\
airquality & comparison & CO$_2$ & CO$_2$ & CO$_2$+\allowbreak{}OAF+\allowbreak{}Occ \\
airquality & verification, diagnosis, explanation & CO$_2$ & CO$_2$+\allowbreak{}Occ & CO$_2$+\allowbreak{}OAF+\allowbreak{}Occ \\
equipment & status, summary & Pwr & Pwr & — \\
equipment & comparison, verification, diagnosis, explanation & Pwr & Pwr & Load+\allowbreak{}Pwr+\allowbreak{}Run \\
occupancy & status, summary, comparison, diagnosis & Occ & Occ & — \\
lighting & status, diagnosis & Lgt & Lgt+\allowbreak{}Occ & — \\
lighting & summary, comparison & Lgt & Lgt & — \\
\bottomrule
\end{tabular}
\par\smallskip\parbox{\textwidth}{\scriptsize 31 question types (domain $\times$ intent) with 76 answer levels; intents with the same levels share a row. —: the question type has fewer levels. Stream codes: T zone air temperature; Sp zone temperature setpoint in effect (heating or cooling); SpC scheduled (occupied/\allowbreak{}unoccupied) setpoint held apart from the setpoint in effect; ActC terminal actuation command (damper or valve command or setpoint, heating or cooling command); ActP terminal actuation response (damper or valve position, terminal airflow); SAT supply air temperature of the air system serving the zone; OAT outside air temperature; Occ occupancy (presence or count); CO$_2$ zone CO$_2$ concentration; OAF outdoor-air flow; Whole whole-building electricity meter; Sub at least one end-use submeter (HVAC, lighting, plug loads, thermal energy); Sub2 two end-use groups metered (the third is the residual of the whole); Pwr equipment input power (electricity or fuel); Run equipment run command or status; Load equipment load (air or water flow, delivered heating or cooling); Lgt lighting power, lighting status, or illuminance.}
 \end{table}

The thermostat-and-meter profile (zone temperature, its setpoint and a whole-building meter) serves
as a reference throughout. Every code edition requires thermostatic control of conditioned zones,
and every building that uses electricity has a utility meter. The profile reaches the first two
answer levels of every thermal question type, since the third needs further streams such as terminal
actuation, and every energy answer level that a whole-building meter alone supports. It reaches none
in the air-quality, equipment, occupancy or lighting questions. Together these are 20 of
the 76 answer levels, an AI readiness of 20/76 = 0.263.

The question set asks for equipment power in every equipment question, so a building whose BAS
records run status, load and supply-air temperature answers none of them. A second scoring, the
HVAC-credit scoring, lets the equipment questions accept those points and is used throughout as a
sensitivity. Under it, run status gives a value alone, and run status with load gives a value with
what it should be compared against. For comparison, verification, diagnosis and explanation, load,
run status and supply-air temperature together give a value with its cause. Each question type keeps
the higher of its answer levels under the two scorings, so no building scores lower. The framework
has no such scoring. It is this study's reading of the question set.

A stock mean does not show which kinds of question buildings can answer, so this study also defines
readiness tiers for three intents: status, summary and diagnosis. The framework defines no tiers. A
building's coverage of an intent is the number of answer levels it reaches across that intent's
question types, divided by the number they offer, so a question type answered at one of its three
levels counts as a third. A building reaches the status, summary or diagnosis tier when its coverage
of that intent reaches one half. Each tier is reached or missed on its own, so a building
can reach the summary or diagnosis tier without the status tier. Like AI readiness, each tier is
computed at the code-required, declared and linked levels.

\subsection{Imputing the unobserved layers}
\label{sec:values}

Three layers are imputed: the code edition, whether each stream is recorded, and the control answers
that EIA itself imputed, which are redrawn from survey buildings that reported them. Description is
not imputed but set as a scenario dimension. The layers are drawn in the order in which they build
on each other, and Appendix~\ref{app:imputation} gives each model and how its parameters are drawn.

\subsubsection{Code edition}
\label{sec:editions}

\begin{figure}
\centering
\includegraphics[width=0.8\textwidth]{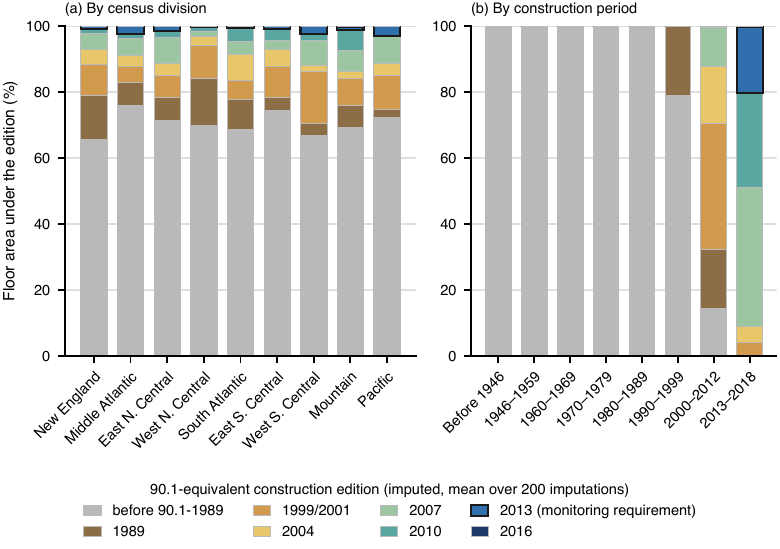}
\caption{Imputed 90.1-equivalent edition in force at construction, as a share of floor area, by
census division (a) and construction period (b).}
\label{fig:fig_editions}
\end{figure}

From the state adoption histories, this study compiles a record of the 90.1-equivalent edition in
force in each state and year, meaning the edition of ASHRAE Standard 90.1 or the state code
equivalent to it. States adopt a 90.1 edition, an IECC edition or a code of their own, and each
adoption is translated to its 90.1 equivalent \citep{athalye2016impacts,tyler2021impacts}. Editions
are matched by overall energy stringency, so a state on an IECC edition or a code of its own is
credited with the sensing requirements of the equivalent 90.1 edition, which its code may not
contain. A sensitivity withholds from such states the credit for the energy-monitoring requirement
that 90.1-2013 introduced. The record holds 291 state-edition rows for 51
jurisdictions (the 50 states and the District of Columbia), summarized by census division in
Appendix~\ref{app:codes}. Of them, 29 are assumed rows, which the federal laboratory
analysis assigns to states with no formal statewide code \citep{belzer2010retrospective}, and a
sensitivity drops them. A building's state and construction year are both imputed. A building's
state is drawn from the states of its census division with probability proportional to each state's
commercial energy use in 2018 \citep{eia2026seds}, as a proxy for the state's share of the
division's commercial buildings, and a sensitivity weights the states equally. Its construction year
is drawn uniformly within its construction period, and its edition is the one in force in that state
and year. A heating and cooling, lighting or electrical system that the survey reports as renovated
since 2000 takes the edition in force in the same state in a year drawn uniformly from 2000, or from
the start of its construction period if later, to 2018. Figure~\ref{fig:fig_editions} shows the
floor-area share imputed to each edition by construction period and census division.

\subsubsection{Code requirements and the code-required level}
\label{sec:clauses}

Each 90.1 edition requires certain data streams to be installed. Table~\ref{tab:clauses} lists the
requirements that enter the estimates, and Appendix~\ref{app:codes} quotes every requirement
reviewed, with its source.

Three streams are present under every edition. Zone temperature and its setpoint come from the
thermostat that every heated or cooled zone must have, and the utility installs the whole-building
electricity meter. Together they form the thermostat-and-meter profile, which sets the code-required
level of almost every building. Before 90.1-2001, and for buildings that predate any code, the
profile is credited from common practice. A building that is neither heated nor cooled lacks the
thermostat streams and scores below it.

Three requirements of 90.1-2013 add streams. Lighting in many more space types must switch off
automatically after occupants leave, which adds occupancy. New buildings of at least
25,000~ft\textsuperscript{2} must monitor any district chilled water, steam or hot water
they receive. Those other than hotels, motels and restaurants must also monitor total, HVAC,
lighting and receptacle electricity, record it every 15 minutes and keep it for
36 months. These buildings are credited with end-use meters, lighting and equipment power.

Two requirements enter only as sensitivities. Editions before 2013 required occupancy sensing only
in space types the survey does not identify. 90.1-2016 requires it for altered lighting, but
replacing lamps or luminaires can comply through scheduled shutoff, and the survey does not separate
such work from other lighting alterations. Seven further requirements enter no estimate. Six apply
under conditions the survey does not record, such as occupant density, climate zone or equipment
capacity, and the seventh requires no recording.

A requirement applies when the edition governing the system it covers (heating and cooling, lighting
or electrical service) is at least the edition that introduced it. Requirements for new construction
follow the construction edition. Requirements that also cover alterations follow the later of the
construction and renovation editions, and requirements for alterations alone apply only to systems
the survey reports as renovated.

\begin{table}
\centering
\footnotesize\linespread{1}\selectfont
\caption{Energy-code requirements that enter the estimates, the data streams each requires and the
edition that introduced it.}
\label{tab:clauses}
\setlength{\tabcolsep}{4pt}
\begin{tabular}{L{3.4cm}L{2.8cm}L{1.6cm}L{2.8cm}L{1.6cm}}
\toprule
Requirement & Data streams & From edition & Applies to & Use \\
\midrule
Zone thermostatic control & Zone temperature, setpoint & All (practice
  before 2001) & Heated or cooled buildings & Code-required \\
Utility revenue meter (not a 90.1 requirement) & Whole-building electricity & All (practice) &
  Buildings using electricity & Code-required \\
Occupancy-sensing lighting control & Occupancy & 90.1-2013 & New lit
  buildings & Code-required \\
Lighting alterations & Occupancy & 90.1-2016 & Renovated lighting & Sensitivity only \\
Electrical energy monitoring & Lighting, equipment power, end-use meters & 90.1-2013 &
  New buildings $\geq$ 25,000~ft\textsuperscript{2} except hotels, motels and restaurants &
  Code-required \\
Non-electric energy monitoring & End-use meters (district chilled water, steam, hot water) &
  90.1-2013 & New buildings $\geq$ 25,000~ft\textsuperscript{2} & Code-required \\
Lighting space control & Occupancy (2004: three space types; 2010: eight space types, where a
  timer switch also complies) & 90.1-2004, 2010 & Lit buildings & Sensitivity only \\
\bottomrule
\end{tabular}
\par\smallskip\parbox{\textwidth}{\footnotesize Use: whether the requirement enters the
code-required level or only a sensitivity.}
\end{table}

The survey does not record whether a building installed what its code required. In 24 new
commercial buildings, the installed controls scored 9.5 of 10 on average for their capability to
meet the code, although only half of the 247 control measures were configured fully compliantly
\citep{rosenberg2017controls}. Neither this nor a measure-by-measure study of compliance
\citep{tyler2021compliance} reports how often building-level monitoring was installed. Installation
of the streams added from 90.1-2013 onward is therefore a scenario dimension, the probability c. At
its reference value, c = 1, the code is credited with everything it requires, so a finding that the
code adds little holds all the more at lower compliance. The sensitivity analysis varies c from 0.5
to 1, because the evidence puts installation well above one half, and tests each conclusion down to
c = 0.05.

\subsubsection{Reported controls and the declared level}
\label{sec:controls}

The reported-controls table maps each control the survey records to the data streams it implies
(Table~\ref{tab:controls}). Each row states the survey question it reads, the streams it declares
and the reasoning behind the mapping, and Appendix~\ref{app:controls} gives every row in full. A BAS that
controls heating or cooling, for example, is taken to read zone temperature, setpoints, supply-air
temperature and the run status of the equipment it commands. Seven rows read equipment the survey
also records, such as an economizer or a chiller plant under BAS control. Each of the seven was
graded strong, moderate to strong, or moderate by how strongly the equipment implies the data
stream, before any effect was computed. Two further rows, graded weak, infer equipment power from a
chiller or boiler plant under BAS control and an end-use meter from a formal energy management plan.
The first is used only in the optimistic scenario described below, and the second only in the
sensitivity that credits reported controls as recorded without a BAS. The table as first written
lacked the rows for demand-controlled ventilation, daylight harvesting and VAV systems, and it read
a BAS that controls lighting as sensing occupancy, where the final table reads it as recording
lighting. That version is kept as a sensitivity. Because it declares no lighting or CO$_2$ stream,
it lowers AI readiness. Each row records that a control exists, and whether its streams are recorded
is the next layer.

\begin{table}
\centering
\footnotesize\linespread{1}\selectfont
\caption{Reported controls, the data streams each is taken to declare, and the estimates that use
them.}
\label{tab:controls}
\setlength{\tabcolsep}{4pt}
\begin{tabular}{L{3.4cm}L{4.3cm}ccc}
\toprule
Reported control & Data streams declared & Primary & Strict & Enriched \\
\midrule
BAS controls heating or cooling & Zone temperature (30), setpoint (23),
  supply-air temperature (40), run status (33) & \checkmark & \checkmark &
  \checkmark \\
Smart thermostat (asked only without a BAS) & Zone temperature, setpoint & \checkmark &
  \checkmark & \checkmark \\
BAS controls heating or cooling, with VAV & Damper or valve command (26) and position
  (17) & \checkmark & \checkmark & \checkmark \\
Lighting occupancy sensors & Occupancy (2) & \checkmark & -- & \checkmark \\
Demand-controlled ventilation & Zone CO$_2$ (1) & \checkmark & -- & \checkmark \\
Daylight harvesting & Lighting (0) & \checkmark & -- & \checkmark \\
BAS controls lighting & Lighting (0) & \checkmark & -- & \checkmark \\
\midrule
Economizer & Outdoor-air temperature (39) & & & \checkmark \\
District chilled water, steam or hot water & Thermal-energy meter (22) & & &
  \checkmark \\
Chillers, BAS controls cooling & Run status and load (28) & & & \checkmark \\
Boilers, BAS controls heating & Run status, load & & & \checkmark \\
Variable-refrigerant-flow system & Zone temperature, setpoint, run status & & &
  \checkmark \\
BAS controls heating or cooling & Scheduled setpoint (1) & & & \checkmark \\
Demand-responsive lighting & Lighting (0) & & & \checkmark \\
\bottomrule
\end{tabular}
\par\smallskip\parbox{\textwidth}{\footnotesize Numbers in parentheses: knowledge graphs in the
reference set, of all 45, that declare the stream (Appendix~\ref{app:controls} counts the 37 with
data). Primary, Strict, Enriched: the primary result and its
two scenarios. \checkmark: row used. --: row dropped by the reference-set test. Scheduling and
dimming rows, which declare no stream, and the two weak rows are omitted. A row without a number
shares the count of the row above or has no counterpart in the reference set.}
\end{table}

\subsubsection{Recording, description and the linked level}
\label{sec:recording}

The survey records neither whether a stream is recorded nor whether it is described, so each stream
is imputed by the route through which it could be recorded (Table~\ref{tab:recording}). Four rules
follow from these routes, and each can remove a stream that the code or the reported controls put in
place. A conventional thermostat keeps no record, while connected thermostats keep their readings
\citep{huchuk2018longitudinal,huchuk2021commercial}, so zone temperature and setpoint count only
through a BAS or a smart thermostat. Monthly bills cannot answer questions about hours or days, so
the whole-building meter counts only where it is an advanced meter, drawn from the state's
commercial advanced-meter share. Because utilities tend to give their largest customers more
sophisticated meters \citep{ferc2006assessment}, a sensitivity credits an interval meter to every
building of at least 25,000~ft\textsuperscript{2}. The streams of other reported controls, such as
occupancy sensors or demand-controlled ventilation, count only where a BAS records them. The BAS
streams in turn count only as far as the building links them. The streams of the 90.1-2013
monitoring requirement count wherever it applies and was installed, because it requires recorded
readings.

\begin{table}
\centering
\footnotesize\linespread{1}\selectfont
\caption{How each group of data streams comes to be recorded, and how that is imputed.}
\label{tab:recording}
\setlength{\tabcolsep}{4pt}
\begin{tabular}{L{4.3cm}L{2.6cm}L{5.8cm}}
\toprule
Data streams & Route & Imputation \\
\midrule
Streams of a BAS (zone temperature, setpoint, supply-air temperature, run status, damper and valve
  position, CO$_2$ and others in Table~\ref{tab:controls}) & BAS trend records & Building drawn as
  linking all, none or part in the reference-set proportions; in a partial building, each stream at
  its reference-set rate \\
Zone temperature and setpoint, no BAS & Smart-thermostat service & Recorded if the building
  reports a smart thermostat \\
Whole-building electricity & Utility advanced meter & Recorded with the state's commercial
  advanced-meter share, 2018; monthly bills not counted \\
End-use electricity, lighting and equipment power & Code-required monitoring &
  Recorded where the requirement applies and was installed \\
Occupancy and lighting & Lighting control through the BAS & Recorded only when the BAS controls
  lighting \\
\bottomrule
\end{tabular}
\par\smallskip\parbox{\textwidth}{\footnotesize Sources: CBECS 2018 for the BAS, smart-thermostat
and lighting-control answers \citep{eia2022cbecsmicrodata}; Form EIA-861 for advanced meters
\citep{eia2018form861}; the monitoring requirement of Standard 90.1-2013 for monitoring; the
knowledge graphs of the reference set for linkage, with how many link all, none or part of their
streams in Appendix~\ref{app:controls}.}
\end{table}

Linkage is drawn from the reference set. Of its 37 knowledge graphs whose declared streams
answer at least one question, 22 link every declared stream to recorded series, 9
link none, and 6 link part, keeping 0.30 to 0.75 of their declared AI readiness.
Appendix~\ref{app:controls} gives these counts by stream. The imputation keeps this all-or-nothing
pattern. A building with a BAS is drawn as linking all, none or part of its BAS streams in these
proportions, and in a building drawn as partial each stream is linked at its rate among the partial
knowledge graphs. Seven of the nine that link none publish no time-series reference at
all, which may reflect how the graphs were released, so a sensitivity treats them as missing. The
knowledge graphs are taken as accurate, although they can leave out links that exist or link a
mislabeled series, so the transfer to the stock is tested in both directions.

A recorded stream must also be described, and no national survey or disclosure program records
whether it is. For streams recorded outside a BAS, description is a scenario dimension, the
probability d, set to 1 in the primary result and searched down to 0.05 because no data narrow it.
For BAS streams, the reference set's linkage already includes description, because a stream counts
as linked there only where the knowledge graph ties it to the equipment or space it measures. This
credits a surveyed BAS with the description of buildings whose owners published a knowledge graph,
which many BAS lack because their point names are inconsistent \citep{kramer2020business}, so it
errs upward, and a scenario draws BAS streams against d as well. A building at the
thermostat-and-meter profile thus has a code-required level of 0.263 but a linked level of
0 if its thermostat is conventional and its meter is not an advanced meter.

\subsection{Primary result and scenarios}
\label{sec:estimates}

The primary result applies the imputation models to the rows of the reported-controls table marked
Primary in Table~\ref{tab:controls}, with c = d = 1. Two scenarios bracket it, a strict one below
and an enriched one above, and an optimistic scenario combines every upward assumption tested.

The strict scenario tests the reported-controls table against real buildings. The table assumes, for
example, that a BAS that controls lighting holds lighting data. If almost none of the reference
set's knowledge graphs list a kind of stream, that assumption has little support. The strict
scenario therefore counts a stream recorded through a BAS only if at least a tenth of the
45 knowledge graphs, that is, five or more, list its class. The choice of a tenth does not
matter, because every class is listed in at least 17 of the 45 knowledge graphs or in at
most 2 (Table~\ref{tab:controls}: occupancy in 2, CO$_2$ and the scheduled setpoint in 1 each,
lighting in none), so any threshold from 3 to 17 drops the same four classes. The test
applies only to streams recorded through a BAS, so the whole-building meter stays, because its
recording comes from utility statistics. The knowledge graphs describe HVAC points far more often
than lighting, so the rarity of lighting and occupancy streams may partly reflect what their authors
chose to describe. CO$_2$ and the scheduled setpoint are ventilation and control points and carry no
such caveat.

The enriched scenario adds the seven rows that read equipment the survey records, in the lower block
of Table~\ref{tab:controls}. The optimistic scenario combines the enriched rows (with c = d = 1, as
in the primary result), the reference set's odds of linkage multiplied by about 20 (a shift of
$\delta = 3$ on the log-odds scale), the weak row that reads a BAS-controlled chiller or boiler
plant as metering equipment power, lighting occupancy sensors recorded without a BAS, and an
interval meter in every building of at least 25,000~ft\textsuperscript{2}. Each scenario is imputed
in the same way as the primary result.

\subsection{Weighting and multiple imputation}
\label{sec:weighting}

Floor area is the primary unit of every stock statement. Each record counts by its final weight
times its floor area. Two further weightings answer different questions. Weighting by each
building's annual energy use gives the share of commercial energy used in buildings at each tier,
and the final weight alone gives shares of buildings. Results are reported by building type,
construction period, size and census division. The size classes of at most 5,000,
5,001--24,999 and at least 25,000 ft\textsuperscript{2} follow the floor-area
thresholds of the code requirements.

The analysis creates 200 imputations of the survey, each in three steps. First, it redraws
the linkage model by resampling the reference set's knowledge graphs with replacement and
re-estimating the proportions that link all, none or part of their streams, and each stream's rate
in the partly linked ones. A stream that fewer than five partly linked knowledge graphs list takes a
uniform prior updated by those few, so that a rate resting on one or two knowledge graphs is not
treated as known. Redrawing these parameters in every imputation carries their uncertainty into the
result \citep{rubin1987multiple}. Second, it replaces each control answer that EIA itself imputed,
which the file flags \citep{eia2023cbecsguide}, with the answer of a randomly drawn building of the
same type, size class and construction period that answered the question itself, or of a coarser
group where fewer than ten such buildings did. Third, it draws the unobserved layers as described
above, computes each building's three levels and tiers, and computes each stock estimate with its
sampling variance from the survey's jackknife over the 151 replicate weights.

Rubin's rules then combine the estimates. The point estimate is their mean. Its variance is the mean
sampling variance plus (1 + 1/M) times the variance between imputations, where M is the number of
imputations. Intervals are 95\% intervals with the degrees of freedom of \citet{barnard1999small}.
An estimate's fraction of missing information is the share of its variance that comes from the
imputed layers. The number of imputations follows the rule of \citet{vonhippel2020imputations},
which asks for $1 + 200F^2$ imputations for a fraction of missing information $F$, so that the
standard errors would vary by no more than 5\% if the imputation were repeated. The largest fraction
here is 0.88, for which the rule asks for fewer than the 200 used.

\subsection{Analyses and sensitivities}
\label{sec:inference}

Each hypothesis is answered with estimates of the primary result and their intervals
(Table~\ref{tab:design}). H1, that most floor area answers less than half of the status questions,
is answered by the share of floor area below the status tier. H2, that AI readiness rises with size,
is answered by the mean linked level of each size class and by the ratio of the status tier's share
of floor area to its share of buildings, which exceeds 1 when the tier lies mostly in large
buildings. H3, that the energy code explains little of the stock's AI readiness, is answered by the
share of floor area at the thermostat-and-meter profile and by the code-required level of each
construction period and edition. As indirect evidence for H3, the linked level is regressed on
construction period, building type, size class and reported controls, weighted by floor area, and a
Shapley decomposition splits the $R^2$ among the four by averaging each factor's added $R^2$ over
every order of entry. A period share no larger than the type's is consistent with H3 but cannot
confirm it, because construction period also tracks the spread of controls.

Four further analyses describe where AI readiness comes from and what would raise it. First, a
second Shapley decomposition splits each building's linked level among four groups of data streams:
the thermostat, metering, the HVAC points a BAS reads, and occupancy, lighting and CO$_2$. The
shares sum exactly to the building's value. Second, the same method splits the drop from the
declared to the linked level among the four recording rules. Third, one data stream, or the bundle
that the 90.1-2013 monitoring requirement asks for, is added to every building, and the bundle also
to only the buildings the requirement would cover if they were new. Each addition is scored by its
gain in the floor-area shares at the summary and diagnosis tiers. Fourth, because tier shares
depend on the cutoff of one half, the share of floor area at or above every level of AI readiness
is reported, and the cutoff is varied from 0.4 to 0.6.

Three analyses test how far the conclusions depend on what the data cannot settle. First, the
compliance c and the description d are lowered from 1 to 0.05, and the value at which each
conclusion would change, if any, is reported. Second, the reference set's proportions and rates of
linkage are shifted by $\delta$ on the log-odds scale, from $-3$ to 3, which multiplies the odds of
linkage by 0.05 to 20. The shift tests the transfer to the stock in both directions, because the
reference set may overstate linkage, leaning toward university campuses and owners who built a
knowledge graph, or understate it, through links its authors left out. The tipping point is the
shift at which a conclusion's interval first includes the opposite conclusion
\citep{liublinska2014sensitivity}. Third, a variance-based sensitivity analysis
\citep{sobol2001global,saltelli2010variance} splits the variation in the stock mean and the tier
shares among c, d and $\delta$, with c from 0.5 to 1 and d first from 0.5 to 1 and then from 0.05
to 1, because the split depends on the ranges searched. Inputs estimated from data stay out of this
split, because their uncertainty is already in the variance between imputations. The remaining
tests in Table~\ref{tab:inputs} each vary one premise at a time.

\subsection{Checks and reproducibility}
\label{sec:checks}

Three checks tie the inputs to published values. The final weights reproduce the stock totals that
EIA publishes in its Table B1, 5,918 thousand buildings, as published, and 96,527
million ft\textsuperscript{2} of floor area against the published 96,528 million. The
difference comes from the public file's top-coding of floor area at 2.1 million
ft\textsuperscript{2} for 20 records. The jackknife implementation reproduces the worked
example in EIA's user's guide, the number of buildings of 1,001 to 5,000~ft\textsuperscript{2},
2,832,884, with a standard error of 139,667. The question set as implemented here
returns the framework's published values of 7, 12, 20, 49 and
37 of 76 answer levels for its minimal, comfort and standard profiles, its metered offices
and its small offices, and it recomputes all 90 values of the reference set, the declared
and linked levels of its 45 knowledge graphs. The framework's standard profile is this study's
thermostat-and-meter profile.

The imputation model is checked against the reference set it is fitted to. The reference set's
counts of knowledge graphs that link all, none or part of their data, and each stream's declared and
linked counts, should fall within the ranges simulated from the fitted model. A back-cast, which
would run the method on the buildings of the reference set from their survey attributes and compare
the predicted with the measured levels, was not possible, because none of the buildings of the
reference set has a public building type or construction period, although 35 of its
knowledge graphs give floor area.

Every input file is checked against a recorded checksum, and every random draw starts from a fixed
seed, so each stage reproduces exactly. Each mapping applied to the inputs is a row in
Tables~\ref{tab:clauses}--\ref{tab:recording} or in the edition record. The table of reported
numbers is regenerated from the stage outputs, so each number in the paper traces to the file and
stage that produced it. %

\section{Results}
\label{sec:results}

\subsection{Level of AI readiness}
\label{sec:res-level}

Weighted by floor area, the 2018 U.S. commercial stock reaches a mean linked level of 0.145
[0.127--0.163] (Table~\ref{tab:table_headline}). The fraction of missing information of
the primary mean is 0.85, so most of its uncertainty comes from the layers the survey does
not record and little from sampling. Counted by building, the mean linked level is 0.089, with
41.4\% of buildings holding no linked data. For scale, the thermostat-and-meter profile
scores 0.263, and the framework's metered-office profile 0.645, 49 of the 76
answer levels. By domain, the floor area that reaches the first level of at least one question type
is 52.5\% for energy, 37.1\% for thermal conditions, 11.7\% for lighting, 8.2\% for occupancy, 4.9\%
for air quality and 1.1\% for equipment.

\begin{table}
\centering
\footnotesize\linespread{1}\selectfont
\caption{AI readiness and hypothesis estimates for the 2018 U.S. commercial stock.}
\label{tab:table_headline}
\setlength{\tabcolsep}{4pt}
\begin{tabular}{L{3.6cm}L{2.0cm}L{2.1cm}L{2.0cm}L{2.0cm}}
\toprule
Quantity & Strict (low)\textsuperscript{b} & Primary & Enriched (high) & Missing\newline information\textsuperscript{a} \\
\midrule
Code-required level (mean) & 0.258\newline [0.255, 0.261] & 0.258\newline [0.255, 0.261] & 0.258\newline [0.255, 0.261] & 0.25 \\
Declared level (mean) & 0.344\newline [0.337, 0.351] & 0.344\newline [0.337, 0.351] & 0.353\newline [0.346, 0.360] & 0.04 \\
Linked level (mean) & 0.122\newline [0.109, 0.135] & 0.145\newline [0.127, 0.163] & 0.151\newline [0.131, 0.171] & 0.85 \\
Status tier (\% of floor area) & 0.5\newline [0.3, 1.1] & 8.0\newline [5.9, 10.6] & 8.1\newline [6.0, 11.0] & 0.77 \\
Summary tier (\%) & 1.1\newline [0.6, 1.9] & 8.7\newline [6.6, 11.3] & 8.8\newline [6.7, 11.6] & 0.75 \\
Diagnosis tier (\%) & 0.0\newline [0.0, 0.0] & 4.9\newline [3.6, 6.8] & 5.2\newline [3.8, 7.2] & 0.72 \\
No linked data (\%) & 29.8\newline [26.0, 33.9] & 30.0\newline [26.4, 34.0] & 29.7\newline [25.9, 33.8] & 0.75 \\
Linked level (mean), energy-weighted & 0.135\newline [0.118, 0.151] & 0.163\newline [0.141, 0.185] & 0.171\newline [0.146, 0.196] & 0.88 \\
Linked level (mean), building-weighted & 0.083\newline [0.075, 0.090] & 0.089\newline [0.081, 0.097] & 0.091\newline [0.082, 0.099] & 0.58 \\
\textit{Hypotheses} &  &  &  &  \\
H1: below the status tier (\%) & 99.5\newline [98.9, 99.7] & 92.0\newline [89.4, 94.1] & 91.9\newline [89.0, 94.0] & 0.77 \\
H2: status tier, floor-area share $\div$ building share\textsuperscript{c} & 6.92\newline [3.39, 14.15] & 4.27\newline [3.10, 5.89] & 4.28\newline [3.14, 5.83] & 0.32 \\
H3: at the thermostat-and-meter profile (\%) & 93.3\newline [92.0, 94.4] & 93.3\newline [92.0, 94.4] & 93.3\newline [92.0, 94.3] & 0.15 \\
H3: Shapley $R^2$ share, construction period $-$ building type\textsuperscript{d} & — & $-$0.023\newline [$-$0.045, $-$0.002] & — & 0.55 \\
\bottomrule
\end{tabular}
\par\smallskip\parbox{\textwidth}{\footnotesize Floor-area weighted unless stated. Brackets: 95\% intervals that combine sampling and imputation uncertainty (multiple imputation, M = 200, Rubin's rules). —: not computed for the scenario.\newline \textsuperscript{a} The fraction of missing information of the primary estimate: the share of its variance due to the imputed code edition, recording and linkage.\newline \textsuperscript{b} Acts at the linked level only. Its code-required and declared levels are the primary's, up to the imputation draw.\newline \textsuperscript{c} H2 holds where the ratio exceeds 1.\newline \textsuperscript{d} At or below zero, construction period explains no more of the variance than building type, which supports H3 without confirming it.}
\end{table}
 
\begin{figure}
\centering
\includegraphics[width=1.0\textwidth]{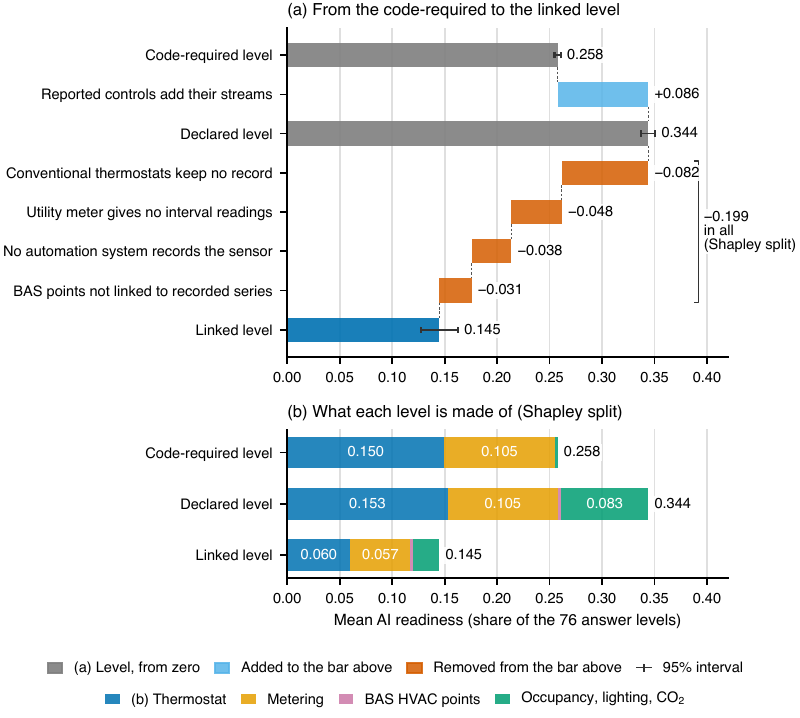}
\caption{Mean AI readiness of the stock from the code-required to the linked level: the change at
each step (a) and the share from each group of data streams (b).}
\label{fig:fig_gap}
\end{figure}

Figure~\ref{fig:fig_gap}(a) traces the stock mean from the code-required to the linked level.
Reported controls add 0.086 to the code-required level, giving a declared level of 0.344,
the stock mean if every stream that codes and reported controls imply were recorded and described.
Of the 0.199 lost between the declared and linked levels, zone temperatures and setpoints read by
conventional thermostats account for 0.082, whole-building meters without interval readings for
0.048, sensors that no automation system records for 0.038, and BAS streams drawn as unlinked for
0.031. The linked level falls below the code-required level for 72.2\% [67.0--77.0] of
floor area and is zero for 30.0\% [26.4--34.0]. The floor area with no linked data lies in
buildings that have neither a smart thermostat nor an imputed advanced meter, and only 17.4\% of it
has a BAS, against 43.0\% of the stock (Appendix~\ref{app:results}).

Of floor area, 92.0\% [89.4--94.1] lies below the status tier, which answers H1. The lower
rows of Table~\ref{tab:table_headline} give the estimate that answers each hypothesis. A thermostat
and a meter answer 3 of the 11 status answer levels and the tier needs 6, so a building
reaches it only with at least two more recorded streams among occupancy, lighting, CO$_2$ and
equipment power. In the primary result these come mostly from a BAS that controls lighting, the one
route that records occupancy and lighting. Few knowledge graphs of the reference set declare them, 2
and 0 of 45, and the strict scenario, which drops them with CO$_2$ and the scheduled
setpoint, puts the status share at 0.5\%. At the code-required level each of the three
tiers holds 1.1\% of floor area, and at the declared level 19 to 20\%, so most of what
reported controls add to the tiers is lost in recording. Moving the tier cutoff from 0.4 to 0.6
lowers the status share from 8.4\% to 4.9\%, so more than 90\% of floor area stays below
the status tier at every cutoff. Appendix~\ref{app:results} gives the share of floor area at
or above every level of AI readiness.

\subsection{Sources and distribution}
\label{sec:res-sources}

Figure~\ref{fig:fig_gap}(b) divides each level among four groups of data streams, by an exact
Shapley decomposition. At the linked level, thermostats supply 0.060, metering 0.057, occupancy,
lighting and CO$_2$ data 0.025, and the HVAC points a BAS reads 0.002. At the
code-required level, thermostats supply 0.150 and metering 0.105, so the recording rules
keep about two-fifths of the thermostat share and about half of the metering share. These shares
divide each level by kind of data, and the losses in panel (a) divide the drop by recording rule, so
the two do not match group for group. The thermostat group, for example, loses 0.093 between the
declared and linked levels, and the thermostat rule removes 0.082.

The BAS's own HVAC points add little because of how the framework's question set is built
\citep{jung2026readiness}. Supply-air temperature, run status and equipment load answer an equipment
or energy question only together with metered energy or equipment power, which neither the survey
nor the reported-controls row for a BAS supplies. Under the HVAC-credit scoring, which credits these
points on their own, the floor area that can answer the simplest equipment questions rises from 1.1
to 31.2\% and the mean linked level from 0.145 to 0.169, but the status tier moves only
from 8.0 to 8.3\%, because the status questions about occupancy, lighting and air quality
remain unanswered for most floor area.

\begin{figure}
\centering
\includegraphics[width=\textwidth]{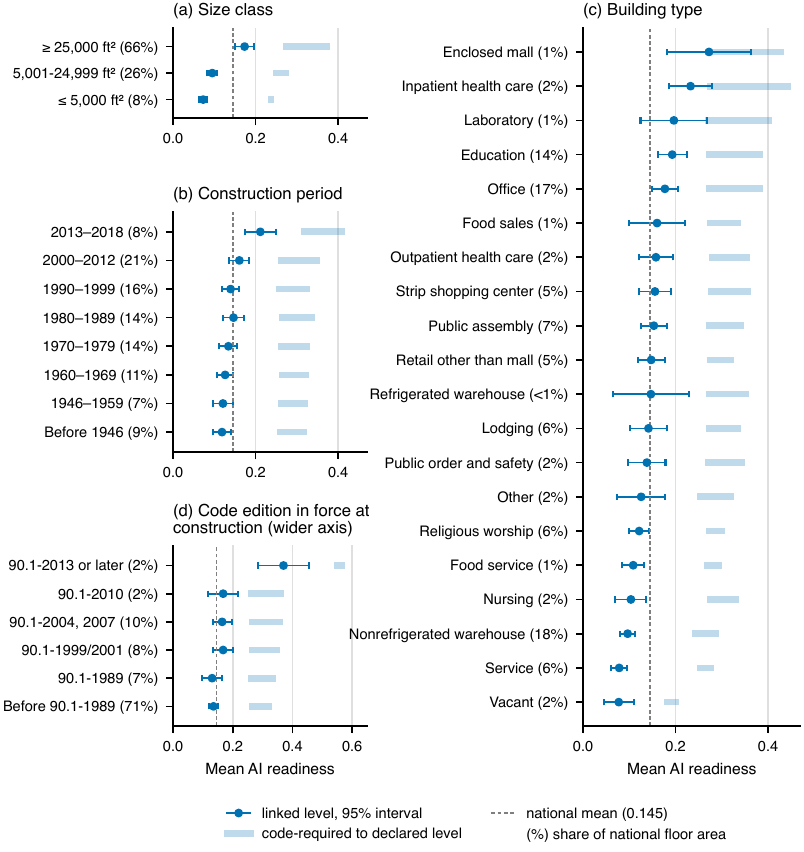}
\caption{Mean linked level by size class (a), construction period (b), building type (c) and code
edition at construction (d).}
\label{fig:fig_segments}
\end{figure}

The linked level rises with size, from 0.073 in buildings of at most 5,000~ft\textsuperscript{2}
through 0.095 to 0.173 at 25,000~ft\textsuperscript{2} or more
(Figure~\ref{fig:fig_segments}). The status tier's share of floor area is 4.27
[3.10--5.89] times its share of buildings, and it holds 9.5\% of commercial
energy use but 1.9\% of buildings. Together these answer H2. Automation follows the same
gradient. A BAS is reported for 9.5\%, 15.5\% and 40.8\% of buildings in the three classes,
and is reported for 10.9 percentage points [6.9--14.9] more of the buildings built in
2000--2018 than of those built before 1960. Small buildings are also the most likely to keep no
linked data at all. No data are linked for 45.4\% of the floor area in buildings of at most
5,000~ft\textsuperscript{2}, 39.5\% in those of 5,001--24,999~ft\textsuperscript{2} and 24.5\% in
those of 25,000~ft\textsuperscript{2} or more, and for 40.2\% of the floor area of
nonrefrigerated warehouses.

The linked level also rises with construction period, from 0.119 for buildings built before 1946 to
0.161 for those built in 2000--2012 and 0.211 for those built in 2013--2018. Among
building types that hold more than 2\% of floor area, inpatient health care (0.232), education
(0.193) and offices (0.177) lead. Nonrefrigerated warehouses, which hold 18.0\%
of the stock's floor area, reach 0.096.

\begin{figure}
\centering
\includegraphics[width=1.0\textwidth]{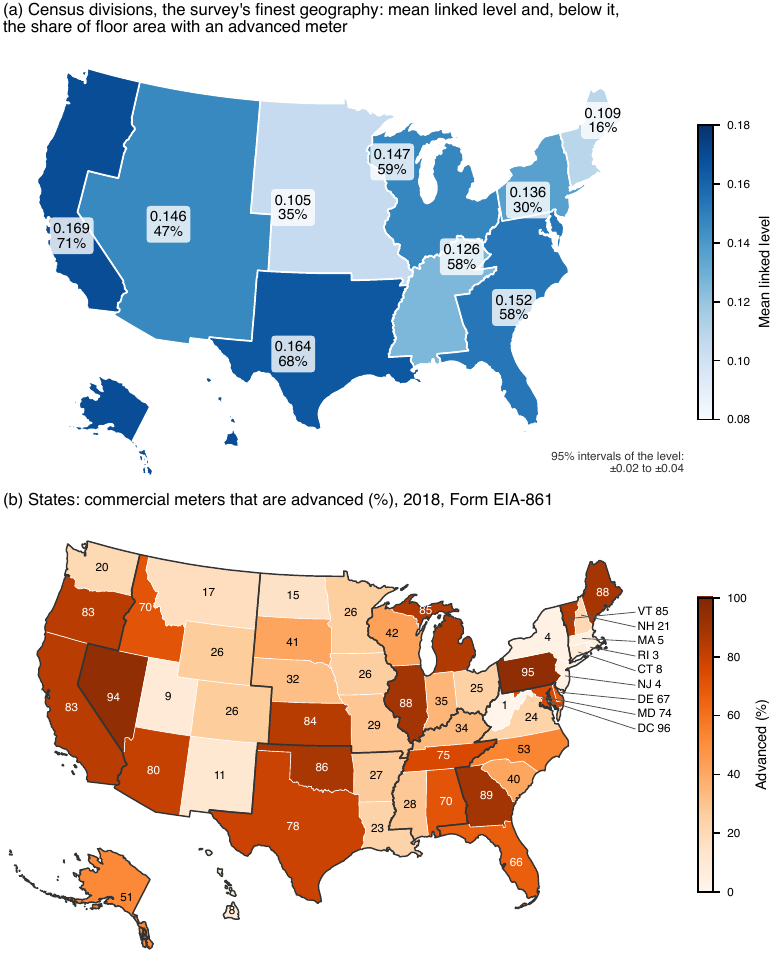}
\caption{Mean linked level by census division (a) and the share of commercial meters that were
advanced in 2018, by state (b).}
\label{fig:fig_geography}
\end{figure}

AI readiness differs across census divisions, from 0.169 in the Pacific division and 0.164 in West
South Central to 0.109 in New England and 0.105 in West North Central
(Figure~\ref{fig:fig_geography}). The ranking follows the utilities' advanced meters more than
automation. The advanced-meter share of floor area is 71.2\% in the Pacific division and 16.2\% in
New England, while the BAS share ranges only from 31.4 to 50.1\% across divisions. Across
the nine divisions, the floor-area-weighted correlation of the linked level is 0.82 with the
advanced-meter share and 0.32 with the BAS share.

\subsection{The energy code}
\label{sec:res-code}

With editions imputed by state and year, 70.7\% of floor area is expected to have been
built before any 90.1-equivalent code and 1.6\% under 90.1-2013 or later
(Figure~\ref{fig:fig_editions}). The code-required level is nearly flat across the stock, with
93.3\% of floor area at the thermostat-and-meter profile, which answers H3. By
construction period, the code-required level hardly changes before 2013, between 0.248 and
0.257, so the rise in the linked level across those periods comes from controls and their
recording. In 2013--2018 it reaches 0.309, as some buildings fall under 90.1-2013.
Construction period's share of the explained variance in the linked level minus building type's is
$-$0.023 [$-$0.045, $-$0.002], which is consistent with H3. Under the HVAC-credit scoring
building type's share is larger by 0.034 [0.014, 0.055].

Figure~\ref{fig:fig_segments}(d) breaks the levels down by the edition in force when each building
was built. From buildings that predate any 90.1-equivalent code through those built under 90.1-2010,
the code-required level stays near 0.25 and the mean linked level between 0.131 and 0.168.
Buildings built under 90.1-2013 or later reach a code-required level of 0.540 and a linked level of
0.370, and 37.9\% of their floor area reaches the status tier. These are also
the newest buildings, and they rest on 79 survey records, so their intervals are wide.
They move the national mean little. Their share falls to 0.8\% and 0.3\% when editions apply one and
two years after adoption, while their linked level stays near 0.37. Without the
monitoring credit in states on an IECC edition or a state code, their linked level is 0.203 and
their status share 19.7\%. Grouped instead by the latest edition among a building's
systems, which counts systems renovated since 2000, 90.1-2013 or later covers 4.0\% of floor area,
with a linked level of 0.246.

\begin{figure}
\centering
\includegraphics[width=\textwidth]{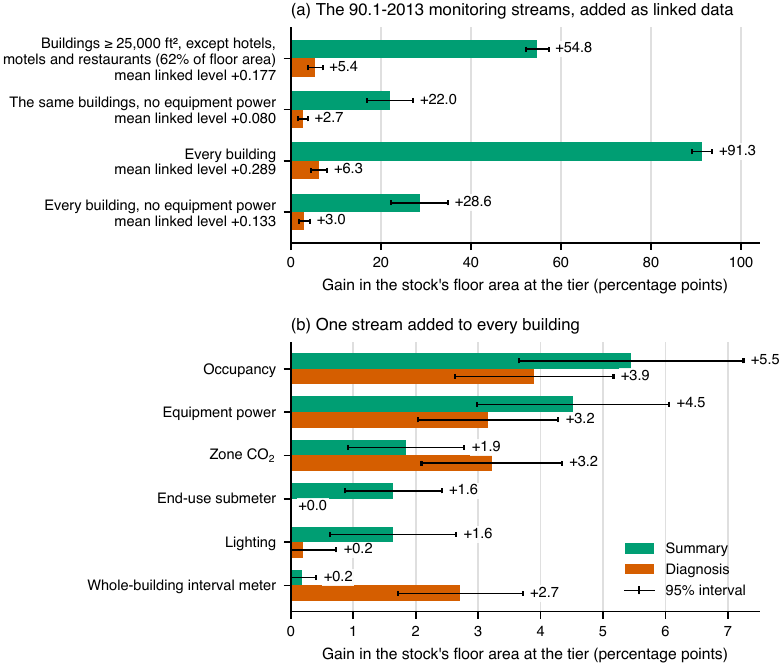}
\caption{Gain in floor area at the summary and diagnosis tiers if the 90.1-2013 monitoring streams
(a) or single data streams that move a tier by 0.5 point or more (b) were added as linked data.}
\label{fig:fig_lever}
\end{figure}

Added to every building to which the 90.1-2013 monitoring requirement would apply if it were new,
those of at least 25,000~ft\textsuperscript{2} other than hotels, motels and restaurants, the
streams it requires raise the mean linked level by 0.177, and the share of the stock's floor area at
the summary and diagnosis tiers by 54.8 and 5.4 percentage points
(Figure~\ref{fig:fig_lever}a). Without equipment power in the bundle, the mean rises by 0.080 and
the summary and diagnosis shares by 22.0 and 2.7 points. Of single streams added to every
building, occupancy and equipment power raise the summary share most, by 5.5 and 4.5
points (Figure~\ref{fig:fig_lever}b). The analysis prices nothing and ranks only what the
monitoring would let the stock answer.

\subsection{Robustness}
\label{sec:res-robust}

\begin{figure}
\centering
\includegraphics[width=0.95\textwidth]{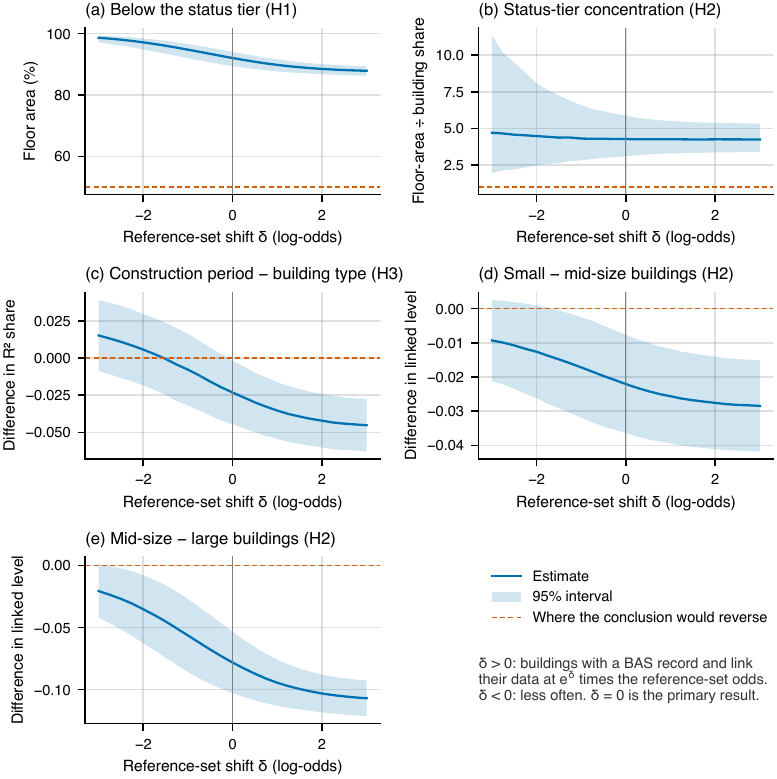}
\caption{The estimates for H1--H3 as the reference set's odds of linkage are shifted by $\delta$ on
the log-odds scale, with the value at which each conclusion would reverse.}
\label{fig:fig_tipping}
\end{figure}

H1 holds at every value tested for the three quantities the data cannot settle. Most floor area
stays below the status tier at every value of compliance and description from 1 down to
0.05, and at every reference-set shift from $\delta = -3$ to 3, which multiplies the odds
of linkage by 0.05 to 20 (Figure~\ref{fig:fig_tipping}). When d also applies to streams
recorded through a BAS, H1 still holds at every value of d, but the level falls fast. At d
= 0.5 the mean linked level is 0.064, and 56.5\% of floor area has no linked data.

H2 holds in part. The status tier stays concentrated in large buildings over the whole range of
compliance, description and $\delta$. The ranking of the three size classes holds
over compliance and description but not over $\delta$. From $\delta = -1.75$ the smallest and the
mid-size buildings can no longer be told apart, as the interval of their difference includes zero,
and from $-3$ neither can the mid-size and the large buildings. When d also applies to BAS
streams, the smallest and the mid-size buildings can no longer be told apart from d = 0.5.
The regression evidence for H3 is the most fragile estimate. Its interval first includes zero at
$\delta = -0.25$, and below $\delta = -1.75$ construction period explains more than building
type. The direct test of H3, the share of floor area at the thermostat-and-meter profile,
does not depend on the reference set.

When the three quantities vary together, the reference-set shift carries most of the variation in
the stock mean, and compliance carries none of it, because few buildings fall under the requirements
it governs. A total-effect index counts a factor's effect alone and together with the
others, so the indices can sum to more than 1. The index of the shift is 0.91 against 0.10 for
description over 0.5 to 1, and 0.73 against 0.28 over description's whole range. The
status and summary tiers rest on streams recorded through a BAS, so for them the shift carries
nearly all of the variation. For the diagnosis tier, description's total effect is 0.06 over the
narrower range and 0.27 over the whole range. Appendix~\ref{app:sensitivities} gives the
full grid of compliance and description and these indices.

\begin{table}
\centering
\footnotesize\linespread{1}\selectfont
\caption{The primary result and the hypothesis estimates under each sensitivity, varied one at a time.}
\label{tab:table_sensitivities}
\setlength{\tabcolsep}{4pt}
\begin{tabular}{L{6.7cm}L{1.02cm}L{1.02cm}L{1.02cm}L{1.02cm}L{1.15cm}}
\toprule
Varied & Mean\newline level & Below\newline status & No\newline data & H2\newline ratio\textsuperscript{a} & H3\newline diff.\textsuperscript{b} \\
\midrule
Primary result & 0.145 & 92.0 & 30.0 & 4.27 & $-$0.023 \\
Question set, domains weighted equally & 0.145 & 92.0 & 30.0 & 4.27 & $-$0.018 \\
Question set, intents weighted equally & 0.143 & 92.0 & 30.0 & 4.27 & $-$0.025 \\
Imputed control answers dropped (contrast) & 0.143 & 92.2 & 30.7 & 4.58 & — \\
States weighted equally within divisions & 0.139 & 92.1 & 33.0 & 4.29 & $-$0.027 \\
No rows assumed for states without a code & 0.145 & 92.0 & 30.0 & 4.27 & $-$0.023 \\
DOE adoption years (17 state--edition pairs) & 0.145 & 92.0 & 30.0 & 4.28 & $-$0.022 \\
Editions applied 1 year after adoption & 0.143 & 92.2 & 30.2 & 4.29 & $-$0.031 \\
Editions applied 2 years after adoption & 0.142 & 92.4 & 30.2 & 4.32 & $-$0.035 \\
Occupancy sensing from 90.1-2004, 2010 & 0.146 & 91.6 & 30.0 & 3.86 & $-$0.021 \\
No monitoring credit in non-90.1 states & 0.142 & 92.3 & 30.2 & 4.23 & $-$0.036 \\
Controls table as first written (no CO$_2$, daylighting or VAV rows, BAS lighting as occupancy) & 0.128 & 99.5 & 30.1 & 6.61 & $-$0.012 \\
Beta(0.5, 0.5) prior for rare streams & 0.146 & 92.0 & 30.0 & 4.29 & $-$0.023 \\
Graphs that link none missing (all 9) & 0.171 & 89.4 & 25.1 & 4.26 & $-$0.044 \\
The same, only the 7 without time series & 0.164 & 90.1 & 26.4 & 4.25 & $-$0.038 \\
Monthly bills credited & 0.193 & 92.0 & 1.6 & 4.27 & $-$0.035 \\
Smart thermostats not credited & 0.136 & 92.1 & 32.6 & 4.27 & $-$0.029 \\
Six controls credited without a BAS & 0.162 & 92.0 & 26.2 & 4.14 & $-$0.028 \\
Advanced-meter shares of 2025 & 0.178 & 92.0 & 10.9 & 4.27 & $-$0.030 \\
Meter only with daily digital access & 0.118 & 92.0 & 45.9 & 4.28 & $-$0.022 \\
Interval meter if $\geq$ 25,000 ft\textsuperscript{2} & 0.177 & 92.0 & 14.0 & 4.28 & $-$0.036 \\
\bottomrule
\end{tabular}
\par\smallskip\parbox{\textwidth}{\footnotesize Point estimates under the full multiple imputation (M = 200, Rubin's rules). Table~\ref{tab:app-E1} gives their 95\% intervals and the tier shares. Mean level: the mean linked level. Below status and No data: the shares of floor area (\%) below the status tier and with no linked data. —: not computed for the contrast.\newline \textsuperscript{a} H2 holds where the concentration ratio exceeds 1. \textsuperscript{b} Construction period minus building type, which supports H3 at or below zero.}
\end{table}
 
No assumption tested lifts most floor area to the status tier. Even the optimistic scenario, which
combines every upward assumption, leaves 74.9\% of floor area below it, and 69.8\% with the
HVAC-credit scoring added. Varied one at a time
(Table~\ref{tab:table_sensitivities}), the assumptions move the mean linked level between 0.118 and
0.193, and how meters are counted matters most. Crediting monthly bills gives the
highest mean. Counting an advanced meter only where the customer has daily digital access to its
readings gives the lowest, with 45.9\% of floor area holding no linked data. Treating the
9 knowledge graphs that link none as missing lowers the share below the status tier the most, to
89.4\%. Weighting each domain or each intent of the question set equally leaves the mean
at 0.145 and 0.143 and the tier shares unchanged. Appendix~\ref{app:sensitivities}
gives each sensitivity with its 95\% interval.

The imputation model reproduces the reference set it was fitted to. All 20 reference-set
statistics checked, among them the numbers of knowledge graphs that link all, none or part of their
streams and the linked count of each stream, fall inside the ranges simulated from the fitted model,
and none has a two-sided predictive p-value below 0.05. The smallest, 0.130, is for the
number of knowledge graphs that link part of their data. The model seldom simulates more than the
observed 6, because most graphs drawn as partly linked end up linking all or none of what their
streams answer. Appendix~\ref{app:imputation} gives every check. Building size, which 35
of the knowledge graphs record, does not predict how a knowledge graph links its data
(likelihood-ratio test, $p = 0.83$), although the test has little power. The control
answers that EIA imputed matter little. Dropping every record with one gives a mean linked level of
0.143.

\section{Discussion}
\label{sec:discussion}

\subsection{AI readiness of the 2018 stock}
\label{sec:disc-support}

Weighted by floor area, the buildings of the 2018 U.S. commercial stock could answer on average
0.145 of the framework's answer levels, a little over half of the 0.263 reached
by a building whose thermostat and meter record zone temperature, its setpoint and whole-building
energy, and 92.0\% of floor area fell below the status tier. By this measure, most of the
stock lacked the recorded data that AI-assisted operation needs, at any value of compliance or
description and when the reference-set odds of linkage are multiplied or divided by
20.

AI readiness in the stock sits on a few steps. Of floor area, 30.0\% records no stream that answers
a question, 32.0\% records only a whole-building interval meter, 6.6\% records only zone temperature
and setpoint, and 7.9\% records both, reaching 0, 8, 12 and 20 of the 76 answer
levels. Together these four steps hold 76.5\% of floor area.
Which step a building reaches depends on three conditions: whether its utility meter records
intervals, whether its zone temperatures are recorded by a connected thermostat or a BAS, and
whether a BAS keeps linked trend records of its other points. Two of the three conditions are
imputed, the meter from the state's advanced-meter share and the BAS linkage from the reference set,
so the estimates describe shares of the stock. They do not give the readiness of any one surveyed
building.

Where other sources measure the same equipment, they agree with the survey or suggest it overstates
automation in small buildings. The survey's BAS share, 43.0\% of floor area, is about the
share reported for 2012 and twice that of 2003 \citep{doe2020sensors}, and above the 33\% of 1999
\citep{brambley2005sensors}. Site visits to commercial buildings in the Pacific Northwest in 2025
found an energy management system in about 1\% of buildings under 25,000~ft\textsuperscript{2}
\citep{neea2026cbsa}, against 9.5\% and 15.5\% that survey respondents reported for its two smaller
size classes in 2018. If the survey overstates automation there, small buildings are less
ready than estimated here, and the gap by size is wider.

Although 43.0\% of floor area has a BAS, only 8.0\% reaches the status tier, and
no published source gives, for a broad population of buildings, the share whose BAS keeps usable
trend records. Analytics programs report that extracting BAS data and interpreting point names are
the main obstacles \citep{kramer2020business}, with a median of 2 hours of in-house set-up per
building for energy information systems and 12 hours for fault detection \citep{lin2022analytics}.
Readiness ratings applied to whole groups of buildings also cluster low. On one UK campus, almost
three quarters of 98 buildings scored 5\% on the smart readiness indicator's minimum-functionality
rating \citep{kourgiozou2023sri}, and the EU study behind the indicator assumed a starting stock
with most buildings in its two lowest classes \citep{eu2020srifinal}.

The stock has moved since 2018. The one change that public data record for existing buildings is the
spread of advanced meters. With the 2025 shares \citep{eia2026form861}, the floor area with no
linked data falls from 30.0 to 10.9\% if every meter's readings reach the
operator, yet 92.0\% of floor area stays below the status tier, because the status
questions ask about spaces and equipment, which a whole-building meter does not measure. No public
source records the controls added to existing buildings since then.

\subsection{Recording as the main loss}
\label{sec:disc-recording}

The stock's equipment holds far more data than the stock keeps. Nearly every building has a
thermostat in each conditioned zone and a utility meter, yet the linked level falls below the
code-required level for 72.2\% of floor area. Three recording rules account for most of
the loss, and each points to a different party. The thermostat is the owner's choice, interval
readings from the meter depend on the utility, and the recording of sensors depends on how a
building's controls are integrated.

Zone temperature and setpoint carry the largest loss, 0.082 of the mean, because a
conventional thermostat acts on the temperature it reads and keeps no record of it. A connected
thermostat keeps that record, and in small commercial buildings such devices already supply data on
how occupants override setpoints \citep{huchuk2021commercial}. In 2018 they added little to the
stock.

The meter carries 0.048. Every electrified building has one, but only an advanced meter
records use at intervals short enough to answer questions about hours or days. In 2018 advanced
meters made up 54.0\% of commercial meters nationally, and between 0.5 and 95.7\%
across states. This loss is shrinking without any action by owners, as advanced meters reached 69\%
of commercial meters by 2021 \citep{eia2023ami}.

Occupancy, lighting and other sensors carry 0.038. A lighting occupancy sensor can switch
its lights with no network connection, so a building that reports one may keep none of its readings,
and these streams count only where an automation system controls lighting. The automation system's
own HVAC points add only 0.002 to the mean. Whether a reported BAS keeps linked data is
the least certain step, and drawing which automation systems link their data removes a further
0.031 from the mean.

\subsection{The reach of the energy code}
\label{sec:disc-code}

The code-required level is nearly flat because the energy code's one requirement to record data
arrived late. Every requirement read here before 90.1-2013 asks for a sensor or a control, and
either can work without keeping a record. Building-level energy monitoring is the first to require
one, and it entered with 90.1-2013. The Department of Energy issued its determination on
that edition in 2014 \citep{doe2014determination}, and states adopted it over the following years.
By 2016, 9 states had adopted 90.1-2013 or an equivalent code, and by 2018, 18,
besides the states that Department of Energy status snapshots place on it (Appendix~\ref{app:codes}).
Only 1.6\% of the 2018 stock's floor area was built under 90.1-2013 or a later edition, or
half that or less if editions took a year or two to reach completed buildings, and
the requirement binds only the part of it in new buildings of at least
25,000~ft\textsuperscript{2}. Most floor area, 93.3\%, sits at the
thermostat-and-meter profile.

Consistent with this, construction period, the survey's clue to the code edition, explains no more
of the variance in AI readiness than building type. The 2013 requirement asks for the
recording the stock lacks, and a code-compliant large office with every series recorded reaches a
ceiling of 0.64 \citep{jung2026readiness}. The requirement reaches the stock only as fast as large
buildings are built, however, and it never reaches small ones.

\subsection{Data, metering and code levers}
\label{sec:disc-levers}

What remains uncertain about the stock's AI readiness lies mostly in how far the reference set's
linkage carries to the stock. No U.S. survey or disclosure program records whether a
building's data are recorded or described. The national survey asks whether a building has an
automation system. It has no question on meter type, interval data or submetering at all.
Asking also whether the system keeps trend records, and whether the building's electricity is
metered at intervals, would replace the two most uncertain layers of this estimate with
observations.

Every building has a meter, and benchmarking and disclosure ordinances already move its monthly
totals to regulators \citep{palmer2016benchmarking,mims2017benchmarking,livingston2018meter}.
Programs that gave operators access to the interval readings of their advanced meters would act on
the existing stock at once. In small buildings, those readings and a connected thermostat are most
of the data within reach.

The code's monitoring requirement binds only new construction and exempts the 33.8\% of
floor area in buildings below 25,000~ft\textsuperscript{2} and the 3.7\% in larger hotels,
motels and restaurants. The recast European directive asks existing non-residential buildings above
a capacity threshold for systems that monitor, log and analyze energy use, where technically and
economically feasible \citep{eu2024epbd}. A U.S. requirement that reached major alterations, or
existing buildings above a size threshold, would act on buildings already standing. Much of the
requirement's gain rests on reading its HVAC monitoring as metering of equipment power. Without that
reading, the summary-tier gain where the requirement applies falls from 54.8 to
22.0 percentage points.

Recorded data are usable only if they can
be matched to the equipment and spaces they measure, and point names vary across vendors and
installers \citep{kramer2020business}. Metadata schemas such as Brick \citep{balaji2016brick} and
the proposed ASHRAE Standard 223P \citep{doe2026ashrae223p} record those relations in a form
software can read. A requirement to describe trend records in such a schema would let software
written for one building use the data of another. None of these steps is priced here. Their costs,
and the energy that better data might save, are left to separate work.

\subsection{Direction of the assumptions}
\label{sec:disc-direction}

The assumptions behind the estimate pull in both directions. Five make the linked level lower than a
more generous reading would: a BAS is credited only with the streams its reported controls imply, a
conventional thermostat records nothing, monthly bills count as no meter data, occupancy and
lighting sensors count only when an automation system controls lighting, and a building's chance of
an interval meter is its state's share of commercial meters, which may understate large buildings.
Crediting an interval meter to every building of at least 25,000~ft\textsuperscript{2} raises it to
0.177. Five others make it higher: compliance and description are held at 1, a surveyed
BAS is credited with the reference set's description, a class of data stream counts whatever its
quality or coverage, and an advanced meter counts as recorded whether or not its readings reach the
operator. Counting only meters whose readings the customer can see lowers the mean linked level to
0.118, and drawing BAS streams against a probability of description of 0.5 lowers it to
0.064. Code editions also apply from the year of adoption, and states on an IECC edition
receive the 90.1 monitoring credit, but their sensitivities change the mean by at most
0.003. The linkage of the reference set could err either way. The strict
choices concern what counts as recorded data, which the question set requires as time series. The
generous ones are premises that no data inform, set at the values most favorable to readiness. The
estimate is therefore strict about recording and generous about everything a record must also be.
Even with the upward assumptions tested here combined, 74.9\% of floor area stays below
the status tier, and 69.8\% under the HVAC-credit scoring, so the main conclusion does not
depend on which group dominates.

\subsection{Limitations}
\label{sec:limitations}

The measure is the readiness framework's \citep{jung2026readiness}. Its question set looks at AI
readiness from one angle, an operator's questions answered from recorded data in six domains. It
leaves out control, optimization and grid services, which need writable setpoints and controllable
loads, and systems outside its domains, such as water, refrigeration and on-site generation. It
weights every answer level equally, so domains with more levels, such as thermal conditions, air
quality and equipment, count for more than occupancy and lighting, and it requires measured energy
for every equipment question, which credits the automation system's HVAC points with little. The
level of AI readiness, the tier shares and the decomposition are therefore specific to this question
set, though not to how it weights its domains or intents, which moves them little. The measure has
not been tied to an outcome such as the energy that analytics save, so a higher linked level means
more operator questions answerable from recorded data and nothing beyond that. Where data are lost,
the code's small reach and the rise with building size rest on what buildings record and which
buildings have an automation system, not on the question set. Having a class of data stream is
treated as enough, whatever its data quality, zone coverage or length of record, so as far as data
quality and coverage go, each value is a ceiling on what a building's data could answer. The
readiness tiers are this study's definition on the framework's question set. Two results
follow largely from definitions. A building reaches the status tier only with recorded space or
equipment streams that, in this model, mostly a BAS that controls lighting supplies, and the share
at the thermostat-and-meter profile follows from the code-requirement table, so H1 and the first
test of H3 measure how much of the stock meets those definitions.

Two imputed layers draw on evidence from outside the surveyed population. The recording of BAS
streams comes from 45 knowledge graphs skewed toward university campuses. The imputation
model reproduces the reference set it was fitted to, but that check does not test
the transfer, and a back-cast was not possible because the buildings of the reference set lack a
public type and construction period. The code edition comes from state adoption records, which miss
local codes that differ from their state's. Linkage is not conditioned on building size, which the
reference set does not predict, and the state is drawn without the survey's climate
region, because the county climate zones behind that region are not public.

Lastly, the survey is a snapshot. Its 2018 cycle is still the latest with released
microdata, and advanced meters have spread since, so the meter loss in this estimate overstates
today's.

\section{Conclusions}
\label{sec:conclusion}

This study estimated how much of the data in the 2018 U.S. commercial building stock could answer
the questions that building operators ask. It imputed each building's code edition and the recording
of its data streams from public evidence, and combined the imputations with Rubin's rules.

Weighted by floor area, the stock could answer on average 0.145 of the framework's answer
levels, or 0.118 to 0.193 when single assumptions change, and 92.0\%
of floor area fell below the status tier. That result holds when compliance, description or the
reference-set odds of linkage are varied. The level itself depends on description. If streams
recorded through a BAS were described with a probability of one half, as other streams can be, the
mean would fall to 0.064. Even when the upward assumptions tested here are combined, three
quarters of floor area stays below the status tier. The status tier itself holds
0.5\% of floor area in the strict scenario and 8.1\% in the enriched one,
because it rests on occupancy and lighting streams that few knowledge graphs of the reference set
declare. Thermostats and meters supply most of what the stock can answer, and most of what is lost
lies in recording. AI readiness rises with size, from 0.073 in buildings of at most
5,000~ft\textsuperscript{2} to 0.173 in those of 25,000~ft\textsuperscript{2} or more. The
energy code has done little for the 2018 stock, because its one requirement to record data entered
with 90.1-2013 and binds only large new buildings.

The largest gains lie in recording what buildings already measure: the zone temperatures that
thermostats read and discard, and the interval readings that advanced meters take but operators may
not receive. Recording requirements that reach existing buildings, and descriptions that let
recorded data be matched to their equipment, would carry those gains across the stock. A survey item
on trend records and interval metering would replace the most uncertain layers of this estimate with
observations.

Several questions remain open. A sample of buildings with both survey attributes and published
knowledge graphs would permit the back-cast that was not possible here, and so test the transfer
from the reference set to the stock. Assessing the quality, zone coverage and record length of each
stream would turn the ceiling estimated here into an estimate of what each building's data can
answer. A question set for control and grid services would measure readiness for tasks this one
leaves out. How much of each ceiling AI agents realize across the stock is open, as is how well AI
methods work from a whole-building interval meter alone, the only linked data for 32.0\%
of floor area. So is whether buildings with higher AI readiness save more energy through analytics.
The tables of code requirements, reported controls and recording routes can be applied to the next
national survey wherever its questions match those of 2018, to track how the stock changes.

\HardFloatBarrier
\credit{\textbf{Wooyoung Jung:} Conceptualization, Methodology, Software, Validation, Formal analysis,
Investigation, Data curation, Visualization, Writing -- original draft, Writing -- review \&
editing.}
\competing{The author declares no competing interests.}
\availability{The 2018 CBECS public-use microdata are available from the U.S. Energy Information Administration
\citep{eia2022cbecsmicrodata}. The code, assumption tables and results of this study are openly
available at Zenodo (doi:10.5281/zenodo.23966090) and at
\url{https://github.com/humanbuildingsynergy/US-CommercialStock_AI_Readiness}. One command
regenerates every number, figure and table in the paper. Third-party inputs are not redistributed.
A script downloads them from their publishers and checks their SHA-256 checksums. The code is
licensed under Apache-2.0 and the data under CC BY 4.0, except the reference-set tables copied from
the readiness framework \citep{jung2026readiness}, which keep its CC BY-NC 4.0 license.}
\funding{This research did not receive any specific grant from funding agencies in the public, commercial, or
not-for-profit sectors.}
\aiuse{During the preparation of this work the author used Claude (Anthropic) in order to draft and edit
manuscript text. After using this tool, the author reviewed and edited the content as needed and
takes full responsibility for the content of the publication.}

\HardFloatBarrier
\appendix
\counterwithin{table}{section}
\renewcommand{\thetable}{\Alph{section}.\arabic{table}}
\counterwithin{figure}{section}
\renewcommand{\thefigure}{\Alph{section}.\arabic{figure}}

\FloatBarrier
\section{Code editions and code requirements}
\label{app:codes}

The code-required level rests on two records: the edition each state had in force in each year, and
the requirements each edition contains. Table~\ref{tab:app-A1} summarizes the first by census
division, and the state rows with their sources are deposited with the study's data and code.
Table~\ref{tab:app-A2} gives the second in full, each requirement quoted from its source.

\begingroup
\footnotesize\linespread{1}\selectfont
\setlength{\tabcolsep}{4pt}
\begin{longtable}{L{2.7cm}L{1.45cm}L{1.45cm}L{1.45cm}L{1.45cm}L{1.45cm}L{1.6cm}}
\caption{First year any state of each census division had each 90.1 edition in force, and the
division's largest state.}\label{tab:app-A1} \\
\toprule
\textbf{Census division} & \textbf{States} & \textbf{90.1-1989} & \textbf{90.1-2004} & \textbf{90.1-2010} & \textbf{90.1-2013} & \textbf{Largest state (\%)\textsuperscript{a}} \\
\midrule
\endfirsthead
\multicolumn{7}{l}{\textit{Table \thetable{} (continued)}} \\
\toprule
\textbf{Census division} & \textbf{States} & \textbf{90.1-1989} & \textbf{90.1-2004} & \textbf{90.1-2010} & \textbf{90.1-2013} & \textbf{Largest state (\%)\textsuperscript{a}} \\
\midrule
\endhead
\midrule
\multicolumn{7}{r}{\textit{continued on the next page}} \\
\endfoot
\endlastfoot
New England & 6 & 1990 (6) & 2005 (6) & 2012 (4) & 2015 (3) & MA (49) \\
Middle Atlantic & 3 & 1991 (3) & 2007 (3) & 2013 (2) & 2016 (2) & NY (46) \\
East North Central & 5 & 1995 (5) & 2008 (5) & 2013 (4) & 2016 (3) & IL (28) \\
West North Central & 7 & 1992 (4) & 2007 (3) & 2014 (2) & 2016 (1) & MO (30) \\
South Atlantic & 9 & 1995 (9) & 2005 (9) & 2012 (8) & 2015 (2) & FL (27) \\
East South Central & 4 & 2005 (4) & 2007 (4) & 2013 (4) & 2016 (1) & TN (39) \\
West South Central & 4 & 1995 (4) & 2007 (4) & 2015 (2) & 2017 (2) & TX (70) \\
Mountain & 8 & 1995 (8) & 2007 (8) & 2013 (7) & 2016 (3) & AZ (27) \\
Pacific & 5 & 1994 (4) & 2005 (4) & 2016 (4) & 2016 (3) & CA (73) \\
\bottomrule
\end{longtable}
\vspace{-0.5\baselineskip}\noindent\parbox{\textwidth}{\scriptsize First year any state of the division had the edition, a later one or its IECC equivalent in force, with the number of states that had it by 2018 in parentheses, including states placed at that edition or later by a DOE status snapshot.\newline \textsuperscript{a} The state with the most commercial energy use in 2018 (EIA State Energy Data System) and its share of the division, the weight used to draw a building's state.}
\endgroup
\FloatBarrier

\begingroup
\footnotesize\linespread{1}\selectfont
\setlength{\tabcolsep}{4pt}
\LongtableBarrier
\begin{longtable}{lL{2.85cm}L{1.5cm}L{1.0cm}L{1.8cm}L{1.6cm}L{2.2cm}}
\caption{The code-requirement table: each requirement, the streams it names, when it binds and its sources.}\label{tab:app-A2} \\
\toprule
\textbf{Row} & \textbf{Requirement} & \textbf{Streams} & \textbf{From} & \textbf{Condition} & \textbf{Use} & \textbf{Source} \\
\midrule
\endfirsthead
\multicolumn{7}{l}{\textit{Table \thetable{} (continued)}} \\
\toprule
\textbf{Row} & \textbf{Requirement} & \textbf{Streams} & \textbf{From} & \textbf{Condition} & \textbf{Use} & \textbf{Source} \\
\midrule
\endhead
\midrule
\multicolumn{7}{r}{\textit{continued on the next page}} \\
\endfoot
\bottomrule
\endlastfoot
C01 & 90.1 \S{}6.4.3.1.1 zone thermostatic control (2004+\allowbreak{}; \S{}6.2.3.1.1 in 2001) & T+\allowbreak{}Sp & pre-1989 & conditioned space & all results & 90.1-2013 addendum ea; 90.1-2001 addendum k \\
C02 & utility electricity service meter (not a 90.1 requirement) & Whole & pre-1989 & electricity used & all results & Form EIA-861 2018 (commercial meters by state) \\
C03 & 90.1-2004 \S{}9.4.1.2(a) space control (2001 addendum ae) & Occ & 2004 & lighting installed & sensitivity only & 90.1-2001 addendum ae; PNNL-17722; 90.1-2007 addendum x \\
C04 & 90.1-2010 \S{}9.4.1.2(b) space control (2007 addendum x) & Occ & 2010 & lighting installed & sensitivity only & 90.1-2007 addendum x; strike-through read from the render); 90.1-2007 addendum av; 90.1-2010 \\
C05 & 90.1-2013 \S{}9.4.1.1(g)/\allowbreak{}(h) and Table 9.6.1 (2010 addendum by) & Occ & 2013 & lighting installed & all results & DOE determination, 90.1-2013; 90.1-2010 addendum by \\
C06 & 90.1-2016 \S{}9.1.2 lighting alterations (2013 addendum e) & Occ & 2016 & lighting installed & sensitivity only & 90.1-2013 addendum e; 90.1-2010 addendum m; DOE determination, 90.1-2013 \\
C08 & 90.1-2013 \S{}8.4.3 electrical energy monitoring (2010 addenda bz +\allowbreak{} bn) & Lgt+\allowbreak{}Pwr+\allowbreak{}Sub+\allowbreak{}Sub2 & 2013 & electricity used, at least 25,000 ft$^2$, not a hotel, motel or restaurant & all results & DOE determination, 90.1-2013; 90.1-2010 addendum bz; 90.1-2010 addendum bn \\
C11 & 90.1 \S{}6.4.3.8 demand-controlled ventilation & CO$_2$ or Occ & 2004 & densely occupied spaces (not in the survey) & not used (condition not in the survey) & 90.1-2004; 90.1-2004 addendum v; 90.1-2010 addendum bs \\
C12 & 90.1 \S{}6.5.1 air economizer high-limit control & OAT & 2004 & economizer required (not in the survey) & not used (condition not in the survey) & 90.1-2007 addendum cy; 90.1-2010 addendum dw \\
C13 & 90.1-2016 economizer fault detection (2013 addendum el) & OAT+\allowbreak{}SAT & 2016 & economizer required (not in the survey) & not used (condition not in the survey) & 90.1-2013 addendum el \\
C14 & 90.1-2016 \S{}6.1.1.3.1 direct-replacement HVAC (2013 addendum cf) & CO$_2$+\allowbreak{}OAT & 2016 & equipment replaced (not in the survey) & not used (condition not in the survey) & 90.1-2013 addendum cf \\
C15 & 90.1-2013 \S{}10.4.5 whole-building energy monitoring (2010 addendum bn, numbered 10.4.4 there) & Sub (district energy only) & 2013 & at least 25,000 ft$^2$, district chilled water, steam or hot water used & all results & 90.1-2010 addendum bn; DOE determination, 90.1-2013 \\
C16 & 90.1-2013 \S{}6.4.3.10 direct digital control (2010 addendum aa, numbered 6.4.3.11 there; 2013 addendum dy edits the table) & none & 2013 & fan power or plant capacity (not in the survey) & not used & 90.1-2010 addendum aa; 90.1-2013 addendum dy \\
C17 & 90.1-2016 \S{}6.4.3.11 chilled-water plant monitoring (2013 addendum bd) & none & 2016 & chilled-water plant capacity (not in the survey) & not used & 90.1-2013 addendum bd \\
C18 & 90.1-2016 \S{}6.4.3.3.5 hotel and motel guest-room HVAC set-point control (2013 addendum d) & none & 2016 & hotel or motel, more than 50 rooms (not applied) & not used & 90.1-2013 addendum d \\
\end{longtable}
\vspace{-0.5\baselineskip}\noindent\parbox{\textwidth}{\scriptsize Stream codes as in Table~\ref{tab:app-C1}. From: the first 90.1 edition with the requirement (pre-1989: universal practice). Condition: when the requirement binds, in survey terms. Use: all results = the primary result and both scenarios; not used = enters no estimate. Rows C07, C09 and C10, the readiness framework's own reading of 90.1-2022, are not used and are not shown. The full requirement text and the quoted passages are deposited with the data and code.}\par
\endgroup
 
\FloatBarrier
\section{Reported controls and recording}
\label{app:controls}

The declared and linked levels rest on the reported-controls table and on how the reference set
links its data. Table~\ref{tab:app-B1} gives the reasoning behind each row of the table, with the
grade each row received before any effect was computed. Table~\ref{tab:app-B2} gives the linkage
counts from which the recording of BAS streams is imputed.

\begingroup
\footnotesize\linespread{1}\selectfont
\setlength{\tabcolsep}{4pt}
\LongtableBarrier
\begin{longtable}{lL{2.95cm}L{1.5cm}L{1.75cm}L{1.55cm}L{3.4cm}}
\caption{The reported-controls table: each survey question, the streams it declares, how many reference-set knowledge graphs declare them, and the reasoning.}\label{tab:app-B1} \\
\toprule
\textbf{Row} & \textbf{Survey question} & \textbf{Streams} & \textbf{Declaring} & \textbf{Use} & \textbf{Reasoning} \\
\midrule
\endfirsthead
\multicolumn{6}{l}{\textit{Table \thetable{} (continued)}} \\
\toprule
\textbf{Row} & \textbf{Survey question} & \textbf{Streams} & \textbf{Declaring} & \textbf{Use} & \textbf{Reasoning} \\
\midrule
\endhead
\midrule
\multicolumn{6}{r}{\textit{continued on the next page}} \\
\endfoot
\bottomrule
\endlastfoot
E01 & Does the building automation system control the \{heating/\allowbreak{}cooling\} and/\allowbreak{}or lighting? Heating? /\allowbreak{} Cooling? & T+\allowbreak{}Sp+\allowbreak{}SAT+\allowbreak{}Run & 19 of 45 & all three & A BAS that controls heating or cooling reads zone temperature and setpoints, and the supply-air temperature and run status of the systems it commands. Supply-air temperature presumes an air system; hydronic-only buildings would not carry it. \\
E02 & Does the building automation system control ... lighting? & Occ & 2 of 45 & sensitivity only & BAS lighting control taken as occupancy-driven. As a sensitivity only: a BAS that only schedules lighting would carry no occupancy stream; CBECS does not separate the two. \\
E03 & Occupancy sensors that reduce lighting when rooms are unoccupied? & Occ & 2 of 45 & all three & Lighting occupancy sensors are usually local devices with no network point. \\
E04 & Does this building use any `smart' or internet-connected thermostats? & T+\allowbreak{}Sp & 23 of 45 & all three & An internet-connected thermostat exposes zone temperature and setpoint. Asked only of buildings without a BAS, so the row cannot fire in BAS buildings. \\
E05 & Lighting scheduling (lights automatically dimmed or turned off at certain times)? & none & — & all three & A time schedule adds no stream class. \\
E06 & Multi-level lighting or dimming? & none & — & all three & A dimming device adds no stream class. \\
E07 & Demand controlled ventilation (DCV)? & CO$_2$ & 1 of 45 & all three & Demand-controlled ventilation taken as CO$_2$-based (90.1 \S{}6.4.3.8 also allows occupancy-based demand-controlled ventilation). \\
E08 & Daylight harvesting (automatically dims light fixtures when natural light is available)? & Lgt & 0 of 45 & all three & Daylight harvesting needs a photosensor, i.e. illuminance, which the lighting stream includes (`lighting power, lighting status, or illuminance'). \\
E09 & Variable air volume (VAV) system? & ActC+\allowbreak{}ActP & 16 of 45 & all three & BAS-controlled VAV boxes carry damper command and position or airflow. \\
E10 & Does the building automation system control ... lighting? & Lgt & 0 of 45 & all three & A BAS that commands lighting holds the lighting on/\allowbreak{}off status (the lighting stream includes lighting status). \\
E11 & Dedicated outside air system (DOAS)? & none & — & no stream & No stream assumed; a dedicated outdoor-air system need not measure outdoor-air flow. \\
E12 & Economizer cycle? & OAT & 39 of 45 & enriched (strong) & An economizer's high-limit control cannot decide when to use outside air without sensing outdoor-air temperature or enthalpy (90.1 \S{}6.5.1). \\
E13 & District chilled water /\allowbreak{} district steam or hot water used? & Sub & 22 of 45 & enriched (strong) & District energy is sold by metered quantity, so a thermal-energy revenue meter exists, as the electricity utility meter does; the end-use submeter stream includes thermal energy. \\
E14 & Central chillers used for cooling? /\allowbreak{} BAS controls cooling? & Run+\allowbreak{}Load & 28 of 45 & enriched (moderate to strong) & A BAS that commands a chiller plant reads whether it runs and what it delivers (chilled-water temperatures or flow). \\
E15 & Boilers used for heating? /\allowbreak{} BAS controls heating? & Run+\allowbreak{}Load & 28 of 45 & enriched (moderate to strong) & A BAS that commands a heating plant reads whether it runs and what it delivers (hot-water temperatures or flow). \\
E16 & BAS controls heating or cooling? & SpC & 1 of 45 & enriched (moderate) & Occupied/\allowbreak{}unoccupied scheduling is a BAS's most basic function, so a scheduled setpoint exists apart from the setpoint in effect; whether it is kept as its own point varies by system. \\
E17 & Demand responsive lighting (reduces lighting at times of peak demand)? & Lgt & 0 of 45 & enriched (moderate) & Dimming lights on a utility signal requires networked lighting control that knows the lighting state. \\
E18 & VRF system? & T+\allowbreak{}Sp+\allowbreak{}Run & 19 of 45 & enriched (moderate) & A variable-refrigerant-flow (VRF) central controller reports indoor-unit temperature, setpoint and run state (`don't know' counted as no). \\
E19 & Chiller or boiler plant, with a BAS that controls heating or cooling? & Pwr & 21 of 45 & optimistic only (weak) & Chiller or boiler input power is often integrated into the BAS, but not reliably. \\
E20 & Formal energy management plan in which energy targets are set and consistently monitored? & Sub & 22 of 45 & sensitivity only (weak) & A plan with monitored targets suggests some end-use monitoring, but targets can be tracked from utility bills alone. \\
\end{longtable}
\vspace{-0.5\baselineskip}\noindent\parbox{\textwidth}{\scriptsize Declaring: how many of the reference set's 45 knowledge graphs declare every stream of the row. Use: all three = the primary result and the strict and enriched scenarios; the strict scenario removes, from the automation-system routes, every stream that fewer than a tenth of the knowledge graphs declare. In parentheses: the row's grade, how strongly the equipment implies the stream, set before any effect was computed. Reasoning: the row's assumption; the full wording, with its sources, is deposited with the data and code.}\par
\endgroup
 
\begingroup
\footnotesize\linespread{1}\selectfont
\setlength{\tabcolsep}{4pt}
\begin{longtable}{L{4.6cm}rrrL{1.5cm}r}
\caption{How the reference-set knowledge graphs link their declared streams, by knowledge graph
(Panel A) and by stream (Panel B).}\label{tab:app-B2} \\
\toprule
\textbf{Panel A: knowledge graphs} & \textbf{Number\textsuperscript{a}} & \textbf{} & \textbf{} & \textbf{} & \textbf{} \\
\midrule
\endfirsthead
\multicolumn{6}{l}{\textit{Table \thetable{} (continued)}} \\
\midrule
\endhead
\midrule
\multicolumn{6}{r}{\textit{continued on the next page}} \\
\endfoot
\endlastfoot
Link all their declared streams & 22 &  &  &  &  \\
Link none & 9 &  &  &  &  \\
Link part (linked share 0.30--0.75) & 6 &  &  &  &  \\
\midrule
\textbf{Panel B: stream} & \textbf{Declared\textsuperscript{a}} & \textbf{Linked} & \textbf{Ratio} & \textbf{Partial: declared /\allowbreak{} linked} & \textbf{Rate used\textsuperscript{b}} \\
\midrule
T: zone temperature & 30 & 23 & 0.77 & 4 /\allowbreak{} 4 & 0.83 \\
Sp: setpoint in effect & 23 & 17 & 0.74 & 4 /\allowbreak{} 4 & 0.83 \\
SpC: scheduled setpoint & 1 & 0 & 0.00 & 0 /\allowbreak{} 0 & 0.50 \\
ActC: terminal actuation command & 26 & 16 & 0.62 & 3 /\allowbreak{} 3 & 0.80 \\
ActP: terminal actuation response & 17 & 13 & 0.76 & 2 /\allowbreak{} 2 & 0.75 \\
SAT: supply-air temperature & 35 & 28 & 0.80 & 6 /\allowbreak{} 6 & 1.00 \\
OAT: outside-air temperature & 33 & 26 & 0.79 & 6 /\allowbreak{} 6 & 1.00 \\
Occ: occupancy & 2 & 1 & 0.50 & 0 /\allowbreak{} 0 & 0.50 \\
CO$_2$: zone CO$_2$ & 1 & 1 & 1.00 & 1 /\allowbreak{} 1 & 0.67 \\
Whole: whole-building meter & 1 & 0 & 0.00 & 1 /\allowbreak{} 0 & 0.33 \\
Sub: end-use submeter & 22 & 12 & 0.55 & 6 /\allowbreak{} 2 & 0.33 \\
Sub2: two end-use groups metered & 1 & 1 & 1.00 & 1 /\allowbreak{} 1 & 0.67 \\
Pwr: equipment power & 21 & 11 & 0.52 & 5 /\allowbreak{} 1 & 0.20 \\
Run: run status & 30 & 22 & 0.73 & 6 /\allowbreak{} 5 & 0.83 \\
Load: equipment load & 27 & 21 & 0.78 & 6 /\allowbreak{} 6 & 1.00 \\
\bottomrule
\end{longtable}
\vspace{-0.5\baselineskip}\noindent\parbox{\textwidth}{\scriptsize A stream is linked when it carries a time-series reference in the published graph. Stream definitions: Table~\ref{tab:app-C1}. Ratio: linked $\div$ declared.\newline \textsuperscript{a} Counts, in both panels, over the 37 reference-set knowledge graphs whose declared streams answer at least one question (the 45 less 8 whose declared streams answer none).\newline \textsuperscript{b} The stream's linked share among the 6 partly linked graphs, the rate at which a partly linked building links it (Beta(1, 1) posterior mean where fewer than five of them declare the stream; each imputation draws the rate).}
\endgroup

\FloatBarrier
\section{Imputation models and model checks}
\label{app:imputation}

Table~\ref{tab:app-D1} lists the models behind the three imputed layers, and Table~\ref{tab:app-D2}
checks the linkage model against the reference set it was fitted to.

\begingroup
\footnotesize\linespread{1}\selectfont
\setlength{\tabcolsep}{4pt}
\LongtableBarrier
\begin{longtable}{L{2.2cm}L{6.2cm}L{4.4cm}}
\caption{The data source and parameter draw of each imputed layer.}\label{tab:app-D1} \\
\toprule
\textbf{Layer} & \textbf{Data source} & \textbf{Parameter draw} \\
\midrule
\endfirsthead
\multicolumn{3}{l}{\textit{Table \thetable{} (continued)}} \\
\toprule
\textbf{Layer} & \textbf{Data source} & \textbf{Parameter draw} \\
\midrule
\endhead
\midrule
\multicolumn{3}{r}{\textit{continued on the next page}} \\
\endfoot
\bottomrule
\endlastfoot
edition in force & state drawn in proportion to its 2018 commercial energy use within the census division (EIA SEDS); construction year drawn uniformly within the survey's construction period; edition: the newest in force in that state by that year; a renovated system takes the edition of a year drawn from 2000, or its period's start if later, to 2018 & drawn per building per imputation; the state tables are treated as fixed administrative counts (no parameter draw) \\
BAS linkage and recording & the shares of knowledge graphs that link all, none or part of their streams, and each class's linked share among those that declare it, from the 37 reference-set knowledge graphs with data  & each imputation re-estimates these shares on a resample, with replacement, of the 45 knowledge graphs and draws the share of a class that fewer than five declare from Beta(1 +\allowbreak{} linked, 1 +\allowbreak{} unlinked); each building's linkage is then drawn \\
interval meter & 2018 commercial advanced-meter share of the building's drawn state (EIA-861) & drawn per building per imputation; the advanced-meter shares are treated as fixed administrative counts \\
thermostat route & survey: a BAS that controls heating or cooling, or a reported smart thermostat; otherwise none & read from the survey (no imputation) \\
monitoring-requirement recording & code text: 90.1-2013 \S{}8.4.3 records every 15 min and keeps 36 months & read from the code text (no imputation); subject to compliance c \\
EIA-imputed control answers & buildings of the same type, size class and construction period that reported the answer (at least 10, or a coarser group otherwise), for the BAS and its heating, cooling and lighting control, smart thermostat, occupancy sensors, demand-controlled ventilation, VAV, dedicated outdoor-air system and daylight harvesting; when the BAS answer was imputed, its control answers are redrawn with it & redrawn from a donor in every imputation \\
control presence & not imputed: the survey observes the control and the recording layer decides whether its stream counts; the reference-set frequencies define the strict scenario only & — \\
\end{longtable}
\vspace{-0.5\baselineskip}\noindent\parbox{\textwidth}{\scriptsize —: none.}\par
\endgroup
 
\begingroup
\footnotesize\linespread{1}\selectfont
\setlength{\tabcolsep}{4pt}
\begin{longtable}{L{6.0cm}rL{2.6cm}r}
\caption{Reference-set statistics against the 95\% ranges simulated from the fitted linkage model.}\label{tab:app-D2} \\
\toprule
\textbf{Statistic} & \textbf{Observed} & \textbf{Simulated 95\% range} & \textbf{Two-sided p} \\
\midrule
\endfirsthead
\multicolumn{4}{l}{\textit{Table \thetable{} (continued)}} \\
\toprule
\textbf{Statistic} & \textbf{Observed} & \textbf{Simulated 95\% range} & \textbf{Two-sided p} \\
\midrule
\endhead
\midrule
\multicolumn{4}{r}{\textit{continued on the next page}} \\
\endfoot
\endlastfoot
knowledge graphs linking all their data & 22 & 16--31 & 0.63 \\
knowledge graphs linking none & 9 & 4--18 & 0.89 \\
knowledge graphs linking part & 6 & 0--6 & 0.13 \\
smallest linked share, partial & 0.30 & 0.23--0.86 & 0.45 \\
largest linked share, partial & 0.75 & 0.30--0.95 & 0.84 \\
knowledge graphs linking stream T (declared 30) & 23 & 15--28 & 0.96 \\
knowledge graphs linking stream Sp (declared 23) & 17 & 11--22 & 1.00 \\
knowledge graphs linking stream SpC (declared 1) & 0 & 0--1 & 0.62 \\
knowledge graphs linking stream ActC (declared 26) & 16 & 13--24 & 0.39 \\
knowledge graphs linking stream ActP (declared 17) & 13 & 7--16 & 0.93 \\
knowledge graphs linking stream SAT (declared 35) & 28 & 19--32 & 0.80 \\
knowledge graphs linking stream OAT (declared 33) & 26 & 18--31 & 0.89 \\
knowledge graphs linking stream Occ (declared 2) & 1 & 0--2 & 1.00 \\
knowledge graphs linking stream CO$_2$ (declared 1) & 1 & 0--1 & 1.00 \\
knowledge graphs linking stream Whole (declared 1) & 0 & 0--1 & 0.71 \\
knowledge graphs linking stream Sub (declared 22) & 12 & 9--19 & 0.53 \\
knowledge graphs linking stream Sub2 (declared 1) & 1 & 0--1 & 1.00 \\
knowledge graphs linking stream Pwr (declared 21) & 11 & 8--18 & 0.51 \\
knowledge graphs linking stream Run (declared 30) & 22 & 15--27 & 1.00 \\
knowledge graphs linking stream Load (declared 27) & 21 & 14--25 & 0.99 \\
\bottomrule
\end{longtable}
\vspace{-0.5\baselineskip}\noindent\parbox{\textwidth}{\scriptsize 2,000 simulated reference sets from the fitted linkage model, each with its own parameter draw. Every observed value lies inside its simulated range. Two-sided p: twice the smaller of the shares of simulated sets at or above and at or below the observed value, capped at 1.}
\endgroup

\FloatBarrier
\section{Sensitivities and scenario dimensions}
\label{app:sensitivities}

\begin{figure}[H]
\centering
\includegraphics[width=\textwidth]{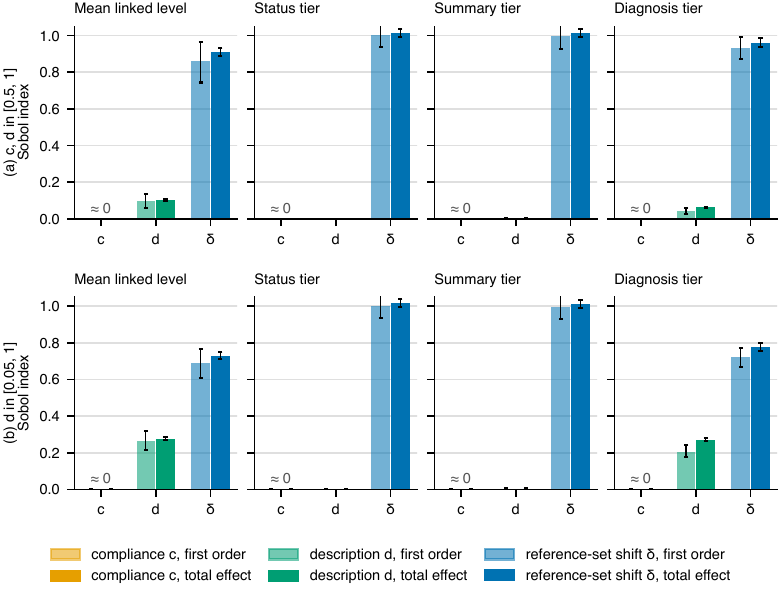}
\caption{Share of each result's variance due to compliance, description and the reference-set shift
(Sobol indices), with description over 0.5--1 and over 0.05--1.}
\label{fig:fig_sobol}
\end{figure}

Table~\ref{tab:app-E1} gives each sensitivity of Table~\ref{tab:table_sensitivities} with its
interval. Table~\ref{tab:app-E2} gives the grid of compliance and description values on which the
status-tier result was tested, and Figure~\ref{fig:fig_sobol} splits the variance among compliance,
description and the reference-set shift.

\begingroup
\footnotesize\linespread{1}\selectfont
\setlength{\tabcolsep}{4pt}
\LongtableBarrier
\begin{longtable}{L{3.0cm}L{2.1cm}L{2.15cm}L{2.15cm}L{2.45cm}}
\caption{AI readiness and tier shares (Panel A) and the hypothesis estimates (Panel B) under each sensitivity, with 95\% intervals.}\label{tab:app-E1} \\
\toprule
\textbf{Panel A} & \textbf{Mean linked level} & \textbf{Status tier (\%)} & \textbf{Summary tier (\%)} & \textbf{Diagnosis tier (\%)} \\
\midrule
\endfirsthead
\multicolumn{5}{l}{\textit{Table \thetable{} (continued)}} \\
\toprule
\endhead
\midrule
\multicolumn{5}{r}{\textit{continued on the next page}} \\
\endfoot
\bottomrule
\endlastfoot
Primary result & \begin{tabular}[t]{@{}l@{}}0.145\\{\scriptsize [0.127, 0.163]}\end{tabular} & \begin{tabular}[t]{@{}l@{}}8.0\\{\scriptsize [5.9, 10.6]}\end{tabular} & \begin{tabular}[t]{@{}l@{}}8.7\\{\scriptsize [6.6, 11.3]}\end{tabular} & \begin{tabular}[t]{@{}l@{}}4.9\\{\scriptsize [3.6, 6.8]}\end{tabular} \\
Each domain of the question set weighted equally & \begin{tabular}[t]{@{}l@{}}0.145\\{\scriptsize [0.127, 0.163]}\end{tabular} & \begin{tabular}[t]{@{}l@{}}8.0\\{\scriptsize [5.9, 10.6]}\end{tabular} & \begin{tabular}[t]{@{}l@{}}8.7\\{\scriptsize [6.6, 11.3]}\end{tabular} & \begin{tabular}[t]{@{}l@{}}4.9\\{\scriptsize [3.6, 6.8]}\end{tabular} \\
Each intent of the question set weighted equally & \begin{tabular}[t]{@{}l@{}}0.143\\{\scriptsize [0.125, 0.161]}\end{tabular} & \begin{tabular}[t]{@{}l@{}}8.0\\{\scriptsize [5.9, 10.6]}\end{tabular} & \begin{tabular}[t]{@{}l@{}}8.7\\{\scriptsize [6.6, 11.3]}\end{tabular} & \begin{tabular}[t]{@{}l@{}}4.9\\{\scriptsize [3.6, 6.8]}\end{tabular} \\
Records with an imputed control answer dropped (contrast) & \begin{tabular}[t]{@{}l@{}}0.143\\{\scriptsize [0.126, 0.161]}\end{tabular} & \begin{tabular}[t]{@{}l@{}}7.8\\{\scriptsize [5.8, 10.5]}\end{tabular} & \begin{tabular}[t]{@{}l@{}}8.5\\{\scriptsize [6.5, 11.2]}\end{tabular} & \begin{tabular}[t]{@{}l@{}}5.0\\{\scriptsize [3.5, 6.9]}\end{tabular} \\
States weighted equally within each division & \begin{tabular}[t]{@{}l@{}}0.139\\{\scriptsize [0.121, 0.157]}\end{tabular} & \begin{tabular}[t]{@{}l@{}}7.9\\{\scriptsize [5.9, 10.5]}\end{tabular} & \begin{tabular}[t]{@{}l@{}}8.5\\{\scriptsize [6.5, 11.1]}\end{tabular} & \begin{tabular}[t]{@{}l@{}}4.6\\{\scriptsize [3.3, 6.4]}\end{tabular} \\
Code-edition table without the rows assumed for states with no formal code & \begin{tabular}[t]{@{}l@{}}0.145\\{\scriptsize [0.127, 0.163]}\end{tabular} & \begin{tabular}[t]{@{}l@{}}8.0\\{\scriptsize [5.9, 10.6]}\end{tabular} & \begin{tabular}[t]{@{}l@{}}8.7\\{\scriptsize [6.6, 11.3]}\end{tabular} & \begin{tabular}[t]{@{}l@{}}4.9\\{\scriptsize [3.6, 6.8]}\end{tabular} \\
DOE's adoption dates for the 17 state-edition pairs where DOE and the federal laboratory analysis disagree & \begin{tabular}[t]{@{}l@{}}0.145\\{\scriptsize [0.127, 0.163]}\end{tabular} & \begin{tabular}[t]{@{}l@{}}8.0\\{\scriptsize [5.9, 10.6]}\end{tabular} & \begin{tabular}[t]{@{}l@{}}8.7\\{\scriptsize [6.7, 11.3]}\end{tabular} & \begin{tabular}[t]{@{}l@{}}4.9\\{\scriptsize [3.6, 6.8]}\end{tabular} \\
Edition in force 1 year before construction or renovation & \begin{tabular}[t]{@{}l@{}}0.143\\{\scriptsize [0.126, 0.161]}\end{tabular} & \begin{tabular}[t]{@{}l@{}}7.8\\{\scriptsize [5.8, 10.4]}\end{tabular} & \begin{tabular}[t]{@{}l@{}}8.3\\{\scriptsize [6.3, 10.8]}\end{tabular} & \begin{tabular}[t]{@{}l@{}}4.9\\{\scriptsize [3.5, 6.7]}\end{tabular} \\
Edition in force 2 years before construction or renovation & \begin{tabular}[t]{@{}l@{}}0.142\\{\scriptsize [0.125, 0.160]}\end{tabular} & \begin{tabular}[t]{@{}l@{}}7.6\\{\scriptsize [5.7, 10.2]}\end{tabular} & \begin{tabular}[t]{@{}l@{}}7.9\\{\scriptsize [6.0, 10.5]}\end{tabular} & \begin{tabular}[t]{@{}l@{}}4.8\\{\scriptsize [3.5, 6.6]}\end{tabular} \\
Occupancy sensing credited from 90.1-2004 and 2010 & \begin{tabular}[t]{@{}l@{}}0.146\\{\scriptsize [0.128, 0.163]}\end{tabular} & \begin{tabular}[t]{@{}l@{}}8.4\\{\scriptsize [6.3, 11.2]}\end{tabular} & \begin{tabular}[t]{@{}l@{}}9.1\\{\scriptsize [7.0, 11.8]}\end{tabular} & \begin{tabular}[t]{@{}l@{}}5.2\\{\scriptsize [3.8, 7.2]}\end{tabular} \\
Occupancy sensing credited for lighting altered under 90.1-2016 & \begin{tabular}[t]{@{}l@{}}0.145\\{\scriptsize [0.127, 0.163]}\end{tabular} & \begin{tabular}[t]{@{}l@{}}8.0\\{\scriptsize [5.9, 10.6]}\end{tabular} & \begin{tabular}[t]{@{}l@{}}8.7\\{\scriptsize [6.6, 11.3]}\end{tabular} & \begin{tabular}[t]{@{}l@{}}4.9\\{\scriptsize [3.6, 6.8]}\end{tabular} \\
No 90.1-2013 monitoring credit where the state's code was IECC or a state code & \begin{tabular}[t]{@{}l@{}}0.142\\{\scriptsize [0.125, 0.160]}\end{tabular} & \begin{tabular}[t]{@{}l@{}}7.7\\{\scriptsize [5.7, 10.3]}\end{tabular} & \begin{tabular}[t]{@{}l@{}}7.9\\{\scriptsize [5.9, 10.5]}\end{tabular} & \begin{tabular}[t]{@{}l@{}}4.8\\{\scriptsize [3.5, 6.7]}\end{tabular} \\
Controls table as first written: no rows for demand-controlled ventilation, daylight harvesting or VAV, and BAS lighting control read as occupancy sensing & \begin{tabular}[t]{@{}l@{}}0.128\\{\scriptsize [0.113, 0.142]}\end{tabular} & \begin{tabular}[t]{@{}l@{}}0.5\\{\scriptsize [0.2, 1.1]}\end{tabular} & \begin{tabular}[t]{@{}l@{}}1.1\\{\scriptsize [0.6, 1.8]}\end{tabular} & \begin{tabular}[t]{@{}l@{}}0.2\\{\scriptsize [0.1, 0.7]}\end{tabular} \\
Rare-stream prior Beta(0.5, 0.5) & \begin{tabular}[t]{@{}l@{}}0.146\\{\scriptsize [0.127, 0.164]}\end{tabular} & \begin{tabular}[t]{@{}l@{}}8.0\\{\scriptsize [5.8, 10.9]}\end{tabular} & \begin{tabular}[t]{@{}l@{}}8.8\\{\scriptsize [6.6, 11.6]}\end{tabular} & \begin{tabular}[t]{@{}l@{}}5.0\\{\scriptsize [3.5, 6.9]}\end{tabular} \\
Knowledge graphs that link none treated as missing (all 9) & \begin{tabular}[t]{@{}l@{}}0.171\\{\scriptsize [0.161, 0.181]}\end{tabular} & \begin{tabular}[t]{@{}l@{}}10.6\\{\scriptsize [8.5, 13.1]}\end{tabular} & \begin{tabular}[t]{@{}l@{}}11.2\\{\scriptsize [9.2, 13.7]}\end{tabular} & \begin{tabular}[t]{@{}l@{}}6.6\\{\scriptsize [5.1, 8.5]}\end{tabular} \\
Knowledge graphs that link none treated as missing (only the 7 with no time-series reference) & \begin{tabular}[t]{@{}l@{}}0.164\\{\scriptsize [0.151, 0.178]}\end{tabular} & \begin{tabular}[t]{@{}l@{}}9.9\\{\scriptsize [7.8, 12.5]}\end{tabular} & \begin{tabular}[t]{@{}l@{}}10.6\\{\scriptsize [8.5, 13.1]}\end{tabular} & \begin{tabular}[t]{@{}l@{}}6.2\\{\scriptsize [4.7, 8.1]}\end{tabular} \\
Monthly bills credited & \begin{tabular}[t]{@{}l@{}}0.193\\{\scriptsize [0.175, 0.211]}\end{tabular} & \begin{tabular}[t]{@{}l@{}}8.0\\{\scriptsize [6.0, 10.6]}\end{tabular} & \begin{tabular}[t]{@{}l@{}}8.9\\{\scriptsize [6.8, 11.5]}\end{tabular} & \begin{tabular}[t]{@{}l@{}}7.7\\{\scriptsize [5.7, 10.2]}\end{tabular} \\
Smart thermostats not credited & \begin{tabular}[t]{@{}l@{}}0.136\\{\scriptsize [0.118, 0.153]}\end{tabular} & \begin{tabular}[t]{@{}l@{}}7.9\\{\scriptsize [5.9, 10.5]}\end{tabular} & \begin{tabular}[t]{@{}l@{}}8.7\\{\scriptsize [6.6, 11.3]}\end{tabular} & \begin{tabular}[t]{@{}l@{}}4.9\\{\scriptsize [3.6, 6.8]}\end{tabular} \\
Six reported controls credited as recorded without a BAS: CO$_2$ sensors for demand-controlled ventilation, economizers, district-energy meters, demand-responsive lighting, VRF systems and energy-target tracking (on the enriched rows, so all six are declared) & \begin{tabular}[t]{@{}l@{}}0.162\\{\scriptsize [0.145, 0.179]}\end{tabular} & \begin{tabular}[t]{@{}l@{}}8.0\\{\scriptsize [6.0, 10.7]}\end{tabular} & \begin{tabular}[t]{@{}l@{}}10.0\\{\scriptsize [7.8, 12.7]}\end{tabular} & \begin{tabular}[t]{@{}l@{}}5.2\\{\scriptsize [3.8, 7.2]}\end{tabular} \\
Advanced meters at 2025 shares (Form EIA-861 2025, latest final year) & \begin{tabular}[t]{@{}l@{}}0.178\\{\scriptsize [0.160, 0.195]}\end{tabular} & \begin{tabular}[t]{@{}l@{}}8.0\\{\scriptsize [6.0, 10.6]}\end{tabular} & \begin{tabular}[t]{@{}l@{}}8.8\\{\scriptsize [6.7, 11.4]}\end{tabular} & \begin{tabular}[t]{@{}l@{}}6.8\\{\scriptsize [5.0, 9.1]}\end{tabular} \\
Whole-building meter recorded with the state's share of commercial customers with daily digital access & \begin{tabular}[t]{@{}l@{}}0.118\\{\scriptsize [0.100, 0.136]}\end{tabular} & \begin{tabular}[t]{@{}l@{}}8.0\\{\scriptsize [5.9, 10.6]}\end{tabular} & \begin{tabular}[t]{@{}l@{}}8.6\\{\scriptsize [6.5, 11.2]}\end{tabular} & \begin{tabular}[t]{@{}l@{}}3.4\\{\scriptsize [2.4, 4.8]}\end{tabular} \\
Interval metering credited for every building of at least 25,000 ft$^2$ & \begin{tabular}[t]{@{}l@{}}0.177\\{\scriptsize [0.159, 0.195]}\end{tabular} & \begin{tabular}[t]{@{}l@{}}8.0\\{\scriptsize [6.0, 10.6]}\end{tabular} & \begin{tabular}[t]{@{}l@{}}8.9\\{\scriptsize [6.8, 11.5]}\end{tabular} & \begin{tabular}[t]{@{}l@{}}7.5\\{\scriptsize [5.5, 10.0]}\end{tabular} \\
\midrule
\textbf{Panel B} & \textbf{H1: below the status tier (\%)} & \textbf{H2: concentration ratio\textsuperscript{a}} & \textbf{Diagnosis tier, declared (\%)\textsuperscript{b}} & \textbf{H3: period $-$ type\textsuperscript{c}} \\
\midrule
Primary result & \begin{tabular}[t]{@{}l@{}}92.0\\{\scriptsize [89.4, 94.1]}\end{tabular} & \begin{tabular}[t]{@{}l@{}}4.273\\{\scriptsize [3.100, 5.888]}\end{tabular} & \begin{tabular}[t]{@{}l@{}}19.3\\{\scriptsize [17.6, 21.1]}\end{tabular} & \begin{tabular}[t]{@{}l@{}}$-$0.023\\{\scriptsize [$-$0.045, $-$0.002]}\end{tabular} \\
Domains weighted equally & \begin{tabular}[t]{@{}l@{}}92.0\\{\scriptsize [89.4, 94.1]}\end{tabular} & \begin{tabular}[t]{@{}l@{}}4.273\\{\scriptsize [3.100, 5.888]}\end{tabular} & \begin{tabular}[t]{@{}l@{}}19.3\\{\scriptsize [17.6, 21.1]}\end{tabular} & \begin{tabular}[t]{@{}l@{}}$-$0.018\\{\scriptsize [$-$0.038, +\allowbreak{}0.001]}\end{tabular} \\
Intents weighted equally & \begin{tabular}[t]{@{}l@{}}92.0\\{\scriptsize [89.4, 94.1]}\end{tabular} & \begin{tabular}[t]{@{}l@{}}4.273\\{\scriptsize [3.100, 5.888]}\end{tabular} & \begin{tabular}[t]{@{}l@{}}19.3\\{\scriptsize [17.6, 21.1]}\end{tabular} & \begin{tabular}[t]{@{}l@{}}$-$0.025\\{\scriptsize [$-$0.046, $-$0.003]}\end{tabular} \\
Imputed answers dropped & \begin{tabular}[t]{@{}l@{}}92.2\\{\scriptsize [89.5, 94.2]}\end{tabular} & \begin{tabular}[t]{@{}l@{}}4.580\\{\scriptsize [3.242, 6.472]}\end{tabular} & \begin{tabular}[t]{@{}l@{}}19.3\\{\scriptsize [17.4, 21.4]}\end{tabular} & — \\
Equal state weights & \begin{tabular}[t]{@{}l@{}}92.1\\{\scriptsize [89.5, 94.1]}\end{tabular} & \begin{tabular}[t]{@{}l@{}}4.287\\{\scriptsize [3.109, 5.910]}\end{tabular} & \begin{tabular}[t]{@{}l@{}}19.2\\{\scriptsize [17.4, 21.0]}\end{tabular} & \begin{tabular}[t]{@{}l@{}}$-$0.027\\{\scriptsize [$-$0.048, $-$0.006]}\end{tabular} \\
No assumed adoption rows & \begin{tabular}[t]{@{}l@{}}92.0\\{\scriptsize [89.4, 94.1]}\end{tabular} & \begin{tabular}[t]{@{}l@{}}4.273\\{\scriptsize [3.100, 5.888]}\end{tabular} & \begin{tabular}[t]{@{}l@{}}19.3\\{\scriptsize [17.6, 21.1]}\end{tabular} & \begin{tabular}[t]{@{}l@{}}$-$0.023\\{\scriptsize [$-$0.045, $-$0.002]}\end{tabular} \\
DOE adoption dates & \begin{tabular}[t]{@{}l@{}}92.0\\{\scriptsize [89.4, 94.1]}\end{tabular} & \begin{tabular}[t]{@{}l@{}}4.276\\{\scriptsize [3.107, 5.886]}\end{tabular} & \begin{tabular}[t]{@{}l@{}}19.3\\{\scriptsize [17.6, 21.2]}\end{tabular} & \begin{tabular}[t]{@{}l@{}}$-$0.022\\{\scriptsize [$-$0.044, $-$0.001]}\end{tabular} \\
Adoption lag, 1 year & \begin{tabular}[t]{@{}l@{}}92.2\\{\scriptsize [89.6, 94.2]}\end{tabular} & \begin{tabular}[t]{@{}l@{}}4.295\\{\scriptsize [3.093, 5.964]}\end{tabular} & \begin{tabular}[t]{@{}l@{}}19.0\\{\scriptsize [17.3, 20.8]}\end{tabular} & \begin{tabular}[t]{@{}l@{}}$-$0.031\\{\scriptsize [$-$0.050, $-$0.011]}\end{tabular} \\
Adoption lag, 2 years & \begin{tabular}[t]{@{}l@{}}92.4\\{\scriptsize [89.8, 94.3]}\end{tabular} & \begin{tabular}[t]{@{}l@{}}4.317\\{\scriptsize [3.092, 6.027]}\end{tabular} & \begin{tabular}[t]{@{}l@{}}18.7\\{\scriptsize [17.0, 20.5]}\end{tabular} & \begin{tabular}[t]{@{}l@{}}$-$0.035\\{\scriptsize [$-$0.054, $-$0.017]}\end{tabular} \\
Occupancy from 2004 and 2010 & \begin{tabular}[t]{@{}l@{}}91.6\\{\scriptsize [88.8, 93.7]}\end{tabular} & \begin{tabular}[t]{@{}l@{}}3.857\\{\scriptsize [2.888, 5.151]}\end{tabular} & \begin{tabular}[t]{@{}l@{}}20.4\\{\scriptsize [18.7, 22.2]}\end{tabular} & \begin{tabular}[t]{@{}l@{}}$-$0.021\\{\scriptsize [$-$0.043, +\allowbreak{}0.000]}\end{tabular} \\
Occupancy from 2016 lighting alterations & \begin{tabular}[t]{@{}l@{}}92.0\\{\scriptsize [89.4, 94.1]}\end{tabular} & \begin{tabular}[t]{@{}l@{}}4.273\\{\scriptsize [3.100, 5.888]}\end{tabular} & \begin{tabular}[t]{@{}l@{}}19.3\\{\scriptsize [17.6, 21.1]}\end{tabular} & \begin{tabular}[t]{@{}l@{}}$-$0.023\\{\scriptsize [$-$0.045, $-$0.002]}\end{tabular} \\
Monitoring credit only under 90.1 & \begin{tabular}[t]{@{}l@{}}92.3\\{\scriptsize [89.7, 94.3]}\end{tabular} & \begin{tabular}[t]{@{}l@{}}4.227\\{\scriptsize [3.041, 5.876]}\end{tabular} & \begin{tabular}[t]{@{}l@{}}18.8\\{\scriptsize [17.1, 20.6]}\end{tabular} & \begin{tabular}[t]{@{}l@{}}$-$0.036\\{\scriptsize [$-$0.054, $-$0.018]}\end{tabular} \\
Controls table as first written & \begin{tabular}[t]{@{}l@{}}99.5\\{\scriptsize [98.9, 99.8]}\end{tabular} & \begin{tabular}[t]{@{}l@{}}6.607\\{\scriptsize [3.314, 13.170]}\end{tabular} & \begin{tabular}[t]{@{}l@{}}1.1\\{\scriptsize [0.6, 1.8]}\end{tabular} & \begin{tabular}[t]{@{}l@{}}$-$0.012\\{\scriptsize [$-$0.037, +\allowbreak{}0.013]}\end{tabular} \\
Jeffreys prior for rare streams & \begin{tabular}[t]{@{}l@{}}92.0\\{\scriptsize [89.1, 94.2]}\end{tabular} & \begin{tabular}[t]{@{}l@{}}4.290\\{\scriptsize [3.137, 5.865]}\end{tabular} & \begin{tabular}[t]{@{}l@{}}19.3\\{\scriptsize [17.6, 21.1]}\end{tabular} & \begin{tabular}[t]{@{}l@{}}$-$0.023\\{\scriptsize [$-$0.046, $-$0.001]}\end{tabular} \\
Link-none graphs missing (all 9) & \begin{tabular}[t]{@{}l@{}}89.4\\{\scriptsize [86.9, 91.5]}\end{tabular} & \begin{tabular}[t]{@{}l@{}}4.262\\{\scriptsize [3.320, 5.471]}\end{tabular} & \begin{tabular}[t]{@{}l@{}}19.3\\{\scriptsize [17.6, 21.1]}\end{tabular} & \begin{tabular}[t]{@{}l@{}}$-$0.044\\{\scriptsize [$-$0.063, $-$0.024]}\end{tabular} \\
Link-none graphs missing (7) & \begin{tabular}[t]{@{}l@{}}90.1\\{\scriptsize [87.5, 92.2]}\end{tabular} & \begin{tabular}[t]{@{}l@{}}4.250\\{\scriptsize [3.269, 5.526]}\end{tabular} & \begin{tabular}[t]{@{}l@{}}19.3\\{\scriptsize [17.6, 21.1]}\end{tabular} & \begin{tabular}[t]{@{}l@{}}$-$0.038\\{\scriptsize [$-$0.058, $-$0.018]}\end{tabular} \\
Monthly bills credited & \begin{tabular}[t]{@{}l@{}}92.0\\{\scriptsize [89.4, 94.0]}\end{tabular} & \begin{tabular}[t]{@{}l@{}}4.273\\{\scriptsize [3.100, 5.889]}\end{tabular} & \begin{tabular}[t]{@{}l@{}}19.3\\{\scriptsize [17.6, 21.1]}\end{tabular} & \begin{tabular}[t]{@{}l@{}}$-$0.035\\{\scriptsize [$-$0.056, $-$0.014]}\end{tabular} \\
No smart thermostats & \begin{tabular}[t]{@{}l@{}}92.1\\{\scriptsize [89.5, 94.1]}\end{tabular} & \begin{tabular}[t]{@{}l@{}}4.271\\{\scriptsize [3.088, 5.906]}\end{tabular} & \begin{tabular}[t]{@{}l@{}}19.3\\{\scriptsize [17.6, 21.1]}\end{tabular} & \begin{tabular}[t]{@{}l@{}}$-$0.029\\{\scriptsize [$-$0.050, $-$0.008]}\end{tabular} \\
Six controls without a BAS & \begin{tabular}[t]{@{}l@{}}92.0\\{\scriptsize [89.3, 94.0]}\end{tabular} & \begin{tabular}[t]{@{}l@{}}4.142\\{\scriptsize [3.035, 5.652]}\end{tabular} & \begin{tabular}[t]{@{}l@{}}19.7\\{\scriptsize [18.0, 21.5]}\end{tabular} & \begin{tabular}[t]{@{}l@{}}$-$0.028\\{\scriptsize [$-$0.050, $-$0.006]}\end{tabular} \\
2025 advanced meters & \begin{tabular}[t]{@{}l@{}}92.0\\{\scriptsize [89.4, 94.0]}\end{tabular} & \begin{tabular}[t]{@{}l@{}}4.273\\{\scriptsize [3.100, 5.889]}\end{tabular} & \begin{tabular}[t]{@{}l@{}}19.3\\{\scriptsize [17.6, 21.1]}\end{tabular} & \begin{tabular}[t]{@{}l@{}}$-$0.030\\{\scriptsize [$-$0.052, $-$0.009]}\end{tabular} \\
Daily digital meter access & \begin{tabular}[t]{@{}l@{}}92.0\\{\scriptsize [89.4, 94.1]}\end{tabular} & \begin{tabular}[t]{@{}l@{}}4.275\\{\scriptsize [3.097, 5.902]}\end{tabular} & \begin{tabular}[t]{@{}l@{}}19.3\\{\scriptsize [17.6, 21.1]}\end{tabular} & \begin{tabular}[t]{@{}l@{}}$-$0.022\\{\scriptsize [$-$0.045, +\allowbreak{}0.001]}\end{tabular} \\
Interval meters, large buildings & \begin{tabular}[t]{@{}l@{}}92.0\\{\scriptsize [89.4, 94.0]}\end{tabular} & \begin{tabular}[t]{@{}l@{}}4.278\\{\scriptsize [3.103, 5.898]}\end{tabular} & \begin{tabular}[t]{@{}l@{}}19.3\\{\scriptsize [17.6, 21.1]}\end{tabular} & \begin{tabular}[t]{@{}l@{}}$-$0.036\\{\scriptsize [$-$0.055, $-$0.016]}\end{tabular} \\
\end{longtable}
\vspace{-0.5\baselineskip}\noindent\parbox{\textwidth}{\scriptsize Primary result and each sensitivity under the full multiple imputation (200 imputations, Rubin's rules, 95\% intervals in brackets). Every interval follows Rubin's rules within its own sensitivity; shares on the logit scale, H2 on the log scale. Dropping imputed answers does not recalibrate the weights. The 2025 advanced-meter shares change only that input; everything else stays at 2018. Daily digital access is commercial customers with daily access $\div$ commercial customers (EIA-861, 2018).\newline \textsuperscript{a} Status-tier share of floor area $\div$ share of buildings.\newline \textsuperscript{b} The diagnosis tier at the declared level.\newline \textsuperscript{c} Construction period minus building type in the Shapley R$^2$ decomposition. At or below zero, construction period explains no more of the variance than building type, which supports H3 without confirming it.}\par
\endgroup

\begingroup
\footnotesize\linespread{1}\selectfont
\setlength{\tabcolsep}{4pt}
\begin{longtable}{L{2.6cm}lL{3.6cm}L{3.6cm}}
\caption{Mean linked level and share of floor area below the status tier over values of compliance
and description.}\label{tab:app-E2} \\
\toprule
\textbf{Panel A: compliance c} & \textbf{Description d} & \textbf{Mean linked level} & \textbf{Below the status tier (\%)} \\
\midrule
\endfirsthead
\multicolumn{4}{l}{\textit{Table \thetable{} (continued)}} \\
\midrule
\endhead
\midrule
\multicolumn{4}{r}{\textit{continued on the next page}} \\
\endfoot
\endlastfoot
0.50 & 0.50 & 0.110 [0.092, 0.127] & 92.4 [89.8, 94.4] \\
0.50 & 0.75 & 0.126 [0.109, 0.144] & 92.3 [89.8, 94.3] \\
0.50 & 1.00 & 0.143 [0.126, 0.161] & 92.2 [89.6, 94.2] \\
0.75 & 0.50 & 0.110 [0.092, 0.128] & 92.4 [89.8, 94.3] \\
0.75 & 0.75 & 0.127 [0.109, 0.144] & 92.3 [89.7, 94.3] \\
0.75 & 1.00 & 0.144 [0.126, 0.162] & 92.1 [89.5, 94.1] \\
1.00 & 0.50 & 0.110 [0.093, 0.128] & 92.3 [89.7, 94.3] \\
1.00 & 0.75 & 0.127 [0.110, 0.145] & 92.2 [89.6, 94.2] \\
1.00 & 1.00 & 0.145 [0.127, 0.163] & 92.0 [89.4, 94.1] \\
\midrule
\textbf{Panel B, over (0, 1]: compliance c} & \textbf{Description d} & \textbf{Mean linked level} & \textbf{Below the status tier (\%)} \\
\midrule
0.05 & 1.00 & 0.142 [0.124, 0.160] & 92.4 [89.8, 94.4] \\
0.10 & 1.00 & 0.142 [0.124, 0.160] & 92.4 [89.8, 94.4] \\
0.25 & 1.00 & 0.143 [0.125, 0.160] & 92.3 [89.8, 94.3] \\
0.50 & 1.00 & 0.143 [0.126, 0.161] & 92.2 [89.6, 94.2] \\
0.75 & 1.00 & 0.144 [0.126, 0.162] & 92.1 [89.5, 94.1] \\
1.00 & 0.05 & 0.081 [0.064, 0.099] & 92.4 [89.8, 94.4] \\
1.00 & 0.10 & 0.084 [0.067, 0.102] & 92.4 [89.8, 94.4] \\
1.00 & 0.25 & 0.094 [0.076, 0.112] & 92.4 [89.8, 94.3] \\
1.00 & 0.50 & 0.110 [0.093, 0.128] & 92.3 [89.7, 94.3] \\
1.00 & 0.75 & 0.127 [0.110, 0.145] & 92.2 [89.6, 94.2] \\
\midrule
\textbf{Panel C, d on BAS streams too: compliance c} & \textbf{Description d} & \textbf{Mean linked level} & \textbf{Below the status tier (\%)} \\
\midrule
1.00 & 0.05 & 0.006 [0.004, 0.007] & 100.0 [$\geq$ 99.99] \\
1.00 & 0.10 & 0.012 [0.010, 0.014] & 100.0 [$\geq$ 99.99] \\
1.00 & 0.25 & 0.030 [0.026, 0.034] & 99.9 [99.6, 100.0] \\
1.00 & 0.50 & 0.064 [0.056, 0.071] & 99.1 [98.5, 99.5] \\
1.00 & 0.75 & 0.102 [0.090, 0.114] & 96.7 [95.3, 97.7] \\
\bottomrule
\end{longtable}
\vspace{-0.5\baselineskip}\noindent\parbox{\textwidth}{\scriptsize Primary result, 200 imputations, Rubin's rules, 95\% intervals in brackets; c = d = 1 is the primary result. Description d applies to data recorded outside the BAS (smart thermostat, advanced meter, monitoring requirement); BAS points already carry it through their reference-set linkage. Panel A: the grid the main analysis reports. Panel B: each dimension alone down to 0.05 with the other at 1; d = 0 would leave nothing described. Panel C: d applied to BAS points as well, with c at 1. Where almost all floor area is below the status tier the interval is given as a bound.}
\endgroup

\FloatBarrier
\section{Further results}
\label{app:results}

\begin{figure}[H]
\centering
\includegraphics[width=0.6\textwidth]{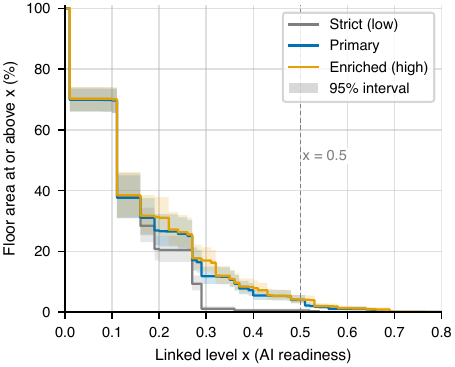}
\caption{Share of floor area at or above each linked level, for the primary result and the strict
and enriched scenarios.}
\label{fig:fig_distribution}
\end{figure}

Figure~\ref{fig:fig_distribution} gives the share of floor area at or above every linked level, the
distribution behind the stock mean. Table~\ref{tab:table_zero_profile} shows where the floor area
with no linked data lies.

\begingroup
\footnotesize\linespread{1}\selectfont
\setlength{\tabcolsep}{4pt}
\begin{longtable}{L{2.4cm}L{3.8cm}L{3.0cm}L{1.8cm}}
\caption{Where the floor area with no linked data lies, by size, type, period and division, against the whole stock.}\label{tab:table_zero_profile} \\
\toprule
& Group & Floor area with no linked data (\%) & Whole stock (\%) \\
\midrule
\endfirsthead
\multicolumn{4}{l}{\textit{Table \thetable{} (continued)}} \\
\toprule
& Group & Floor area with no linked data (\%) & Whole stock (\%) \\
\midrule
\endhead
\midrule
\multicolumn{4}{r}{\textit{continued on the next page}} \\
\endfoot
\endlastfoot
Size class & $\leq$ 5,000 ft\textsuperscript{2} & 12.6 [10.6, 14.9] & 8.3 \\
 & 5,001--24,999 ft\textsuperscript{2} & 33.6 [29.4, 38.0] & 25.5 \\
 & $\geq$ 25,000 ft\textsuperscript{2} & 53.8 [48.6, 59.0] & 66.2 \\
Building type (largest five) & Nonrefrigerated warehouse & 24.1 [19.9, 28.9] & 18.0 \\
 & Office & 14.0 [11.4, 17.1] & 17.2 \\
 & Education & 9.8 [7.3, 13.1] & 14.1 \\
 & Service & 9.4 [7.0, 12.4] & 6.5 \\
 & Public assembly & 7.1 [5.4, 9.2] & 7.4 \\
Construction period & Before 1946 & 11.7 [9.3, 14.6] & 9.5 \\
 & 1946--1959 & 8.2 [6.0, 11.0] & 7.2 \\
 & 1960--1969 & 11.4 [9.2, 14.1] & 10.8 \\
 & 1970--1979 & 14.0 [11.2, 17.5] & 13.5 \\
 & 1980--1989 & 13.4 [10.8, 16.4] & 13.9 \\
 & 1990--1999 & 16.8 [13.1, 21.2] & 16.1 \\
 & 2000--2012 & 18.8 [15.6, 22.4] & 21.3 \\
 & 2013--2018 & 5.8 [4.0, 8.3] & 7.7 \\
Census division & New England & 6.7 [4.7, 9.5] & 3.9 \\
 & Middle Atlantic & 17.1 [13.4, 21.6] & 12.7 \\
 & East North Central & 16.3 [12.0, 21.9] & 18.1 \\
 & West North Central & 13.4 [10.2, 17.4] & 8.8 \\
 & South Atlantic & 16.2 [12.6, 20.6] & 18.6 \\
 & East South Central & 5.6 [3.3, 9.4] & 5.6 \\
 & West South Central & 7.9 [5.6, 11.1] & 12.0 \\
 & Mountain & 8.8 [6.3, 12.2] & 7.9 \\
 & Pacific & 7.9 [5.5, 11.1] & 12.5 \\
Equipment\textsuperscript{a} & BAS & 17.4 [10.0, 28.6] & 43.0 \\
 & Smart thermostat & 0.0 [0.0, 0.0] & 5.8 \\
 & Advanced meter & 0.0 [0.0, 0.0] & 51.5 \\
\bottomrule
\end{longtable}
\vspace{-0.5\baselineskip}\noindent\parbox{\textwidth}{\footnotesize Primary result. Each column gives the distribution of its floor area over the groups. Brackets: 95\% intervals (multiple imputation, M = 200, Rubin's rules).\newline \textsuperscript{a} Not part of the distribution: the share of the column's floor area with a BAS, a smart thermostat or an advanced meter.}
\endgroup

\HardFloatBarrier

\end{document}